\documentclass[a4paper,11pt]{article}
\pdfoutput=1 % if your are submitting a pdflatex (i.e. if you have
\usepackage{jheppub} % for details on the use of the package, please
\usepackage{tikz-cd}
\usepackage{tikz}
\usepackage{tikz-3dplot}
\usepackage{subfigure}

\newcommand{\eps}{\epsilon}

\newcommand{\G}{{\Gamma}}
\newcommand{\Gt}{\tilde{\Gamma}}
\renewcommand{\k}{\kappa}
\renewcommand{\L}{{\mathcal L}}

\newcommand{\La}{{\Lambda}}

\newcommand{\w}{{\omega}}
\newcommand{\wt}{{\tilde{\omega}}}

\newcommand{\C}{\mathbb{C}}
\renewcommand{\H}{\mathbb{H}}
\newcommand{\R}{\mathbb{R}}

\newcommand{\Z}{\mathbb{Z}}
\renewcommand{\S}{\mathcal{S}}

\DeclareMathOperator{\Tr}{Tr}

\DeclareMathOperator{\sgn}{sgn}

\newcommand{\bra}{\langle}
\newcommand{\ket}{\rangle}
\newcommand{\nn}{\nonumber}
\newcommand{\iso}{\cong}

\newcommand{\zv}{{\vec{z}}}
\newcommand{\wv}{{\vec{w}}}

\newcommand{\Qt}{{\tilde{Q}}}

\newcommand{\BB}{\mathbb}

\newcommand{\FR}{\mathfrak}

\newcommand{\bea}{\begin{eqnarray}}
\newcommand{\eea}{\end{eqnarray}}
\newcommand{\sbullet}{\vcenter{\hbox{\tiny$\bullet$}}}

\newcommand{\im}{\operatorname{Im}}

\newcommand{\sdet}{\operatorname{sdet}}
\newcommand{\opn}{\operatorname}
\def\ga{\alpha}
\def\gb{\beta}

\def\Gc{\Gamma}

\def\Gd{\Delta}
\def\ep{\epsilon}
\def\gt{\theta}
\def\gs{\sigma}

\def\gk{\kappa}

\def\Go{\Omega}
\def\go{\omega}
\usepackage{todonotes}

\title{\boldmath 7D supersymmetric Yang-Mills on hypertoric 3-Sasakian manifolds}

\author[a]{Nikolaos Iakovidis,}
\author[a,b]{Jian Qiu}
\author[a]{Andreas Roc\'en,}
\author[b]{Maxim Zabzine}

\affiliation[a]{Department of Mathematics, Uppsala Univeristy,\\Box 480, SE-75106 Uppsala, Sweden}
\affiliation[b]{Department of Physics and Astronomy, Uppsala Univeristy,\\Box 516, SE-75120 Uppsala, Sweden}
\emailAdd{nikolaos.iakovidis@math.uu.se}
\emailAdd{jian.qiu@math.uu.se}
\emailAdd{andreas.rocen@math.uu.se}
\emailAdd{maxim.zabzine@physics.uu.se}

\abstract{We study 7D maximally supersymmetric Yang-Mills theory on 3-Sasakian manifolds. For manifolds whose hyper-K\"ahler cones are hypertoric we derive the perturbative part of the partition function. The answer involves a special function that counts integer lattice points in a rational convex polyhedral cone determined by hypertoric data. This also gives a more geometric structure to previous enumeration results of holomorphic functions in the literature.
Based on physics intuition, we provide a factorisation result for such functions. The full proof of this factorisation using index calculations will be detailed in a forthcoming paper.

}

\begin{document} 
\maketitle
\flushbottom

\section{Introduction} \label{seq:intro}

Localisation of supersymmetric gauge theories on curved manifolds has been extremely fruitful, many exact results have been obtained using this technique for various supersymmetric theories in various dimensions, see \cite{Pestun:2016jze} for a review. The prerequisite for the localisation technique to be applicable is to first formulate a rigid supersymmetric theory on a curve manifold. One systematic approach is to 'rigidify' supergravity theories, which entails freezing the gravity part of sugra at a geometrical background \cite{Festuccia:2011ws} (metric, R-symmetry connection etc) that admits covariantly constant spinors or Killing spinors. Such spinors are then used as parameters for the susy transformation of the rigidified susy gauge theory. The existence of such spinors puts restrictions on the allowed
holonomies of the manifold, see e.g. \cite{Bar, Blau} for the most classical examples. In principle, the sugra approach can provide a classification of such theories, however the equations involved become increasingly complicated as one goes to higher dimensions. For the case of 7D, which we study in this paper, a full classification is beyond us and we instead focus on the most straightforward  background geometry that admits Killing spinors. 

\subsection{The 7D backgrounds}
In 7D the supersymmetric Yang-Mills theory is unique and it is maximally supersymmetric. In \cite{Minahan:2015jta} the theory was placed on the seven-sphere $S^7$ by dimensionally reducing and deforming the 10D Lorentzian version of the theory. In \cite{Polydorou2017} it was argued that the same method could be used for any 7D manifold admitting positive Killing spinors\footnote{Here positive means the scalar curvature of the resulting manifold is positive.}. Such manifolds have been classified by B\"ar \cite{Bar} and fall into the following  types
\begin{enumerate}
\item $S^7$, 16 Killing spinors, 
\item 3-Sasakian manifolds, 3 Killing spinors, 
\item Sasaki-Einstein manifolds, 2 Killing spinors
\item proper $G_2$-manifolds, 1 Killing spinor.
\end{enumerate}
Here the Killing spinors satisfy 
\bea \nabla_{\mu}\eta=+\frac{i}{2}\gamma_{\mu}\eta\label{Killing_spinor_7D}\eea
where $\gamma_{\mu}$ is the 7D gamma matrices.

The next step for the localisation calculation is to organise all fields as 
the de Rham complex of a super-manifold, we call this the \emph{cohomological complex}. On this complex (one particular combination of) susy acts as an equivariant differential. To do this the susy must be realised off-shell. For 7D SYM, the supersymmetry can be taken off-shell for any of the cases listed above, see \cite{Prins:2018hjc, Polydorou2017}, and in \cite{Minahan:2015jta, Polydorou2017} the cohomological complex was written down for the manifolds admitting at least two Killing spinors (Sasaki-Einstein, 3-Sasakian and $S^7$). 
With this complex, the equivariant localisation, applied formally to the path integral, reduces the latter to computing the one-loop approximation round the instanton background. In particular one obtains perturbative partition function
by the one-loop computation round the zero instanton background. 

In order to write the answer in a closed form one typically needs to impose additional symmetry on the manifold. For 7D \emph{toric} Sasaki-Einstein manifolds it was shown in \cite{Polydorou2017} that the perturbative partition function could be written in terms of a generalised quadruple sine function. This is very similar to the result obtained for toric Sasaki-Einstein manifolds in 5D, see \cite{Qiu:2016rev} for a review. A factorisation result was also discussed in \cite{Polydorou2017}, again very similar in spirit to the 5D case in \cite{Qiu:2014oqa}. 

Something that distinguishes the 7D from the 5D case is the possibility of 3-Sasakian structures\footnote{3-Sasakian structures are possible in dimensions $4n-1$. In three dimensions all Sasaki-Einstein manifolds are also 3-Sasakian, so 7D is the lowest dimension in which these two notions are truly distinct.}. In \cite{Polydorou2017} some initial steps were taken towards understanding the role such a 3-Sasakian structure play for localisation calculations on $S^7$, which in addition to being a toric Sasaki-Einstein manifold is also 3-Sasakian. In \cite{Rocen2018} a localisation calculation was performed for a specific 7D 3-Sasakian manifold that is not toric in the Sasaki-Einstein sense. This calculation used that the hyper-K\"ahler cone of this manifold had \emph{hypertoric} symmetry. This paper is a continuation of the work in \cite{Rocen2018} and here we derive a closed form answer for the perturbative partition function for arbitrary 3-Sasakian manifolds whose hyper-K\"ahler cones are hypertoric. The answer is stated in terms of a special function that enumerate integer lattice points in a cone determined by hypertoric data, similar in spirit to how the generalised sine functions count points in cones determined by toric data in the toric Sasaki-Einstein case.

\subsection{Organisation of the paper}
This paper is organised as follows: In sec.\ref{seq:SYM} we briefly review 7D SYM and how to localise it. The main point is that the perturbative partition function can be stated in terms of a superdeterminant and this in turn can be found by considering holomorphic functions on the cone over the manifold. In sec.\ref{sec:hypertoric3S} we recall some facts about hypertoric 3-Sasakian manifolds and how to describe their hypertoric cones by hyperplane arrangements.
In particular we give in sec.\ref{sec_Gctatfp} how to read off the geometry in the neighbourhood of a torus fixed locus.
In sec.\ref{sec:counting} we discuss how to enumerate the holomorphic functions in terms of hyperplanes and formulate this as a count of integer lattice points in a cone. This gives the perturbative partition function in terms hypertoric data.
In sec.\ref{sec:factorisation} we give examples of how to factorise this function as suggested by the geometry read from sec.\ref{sec_Gctatfp}. 
Finally sec.\ref{sec_FitGa} states the main factorisation theorem, 
where we also introduce certain fractional $S_2$ functions as constituent factors of the factorisation. These fractional $S_2$'s reflect the geometry close to the torus fixed locus. We also sketch how to obtain the asymptotic behaviour of our special function, leaving the proofs in a separate paper. 

\section{7D supersymmetric Yang-Mills} \label{seq:SYM}
\subsection{Action and supersymmetry}
Supersymmetric Yang-Mills on compact 7D manifolds can be obtained by dimensionally reducing and deforming the 10-dimensional Lorentzian flat space version of the theory. This approach was taken in \cite{Minahan:2015jta, Polydorou2017} to obtain the following supersymmetric action:
\begin{align}
S_{7D}= \frac{1}{g_{7D}^2} \int d^{7}x \sqrt{-g} \Tr \Big( 
\frac12 F^{MN}F_{MN} 
&- \Psi \G^M D_M \Psi 
+8 \phi^A \phi_A \nn \\
&\quad  +\frac32 \Psi \Lambda \Psi
-2 [\phi^A,\phi^B]\phi^C \varepsilon_{ABC}
\Big)\, .   \label{eq:7Daction}
\end{align}
We use 10D notation with the indices $M,N$ running from $0$ to $9$ and the indices $A,B,C$ over the compactified directions $8,9,0$. 
The fields of the theory consist of the gauge field $A_M$ with field strength $F_{MN}$, the Majorana-Weyl fermion $\Psi$, and the scalars $\phi_A$ coming from the components of $A_M$ along the $8, 9, 0$ directions.
%The $\Gamma^M,\tilde\Gamma^M$ are $16\times 16$ and are components of the 10D %gamma matrices
%\bea \gamma^M=\left[\begin{array}{cc} 0 &\tilde\Gamma^M \\ \Gamma^M & %0\end{array}\right]\nn\eea
%i.e. $\Gamma^M$ maps the plus chirality spinors to the minus chirality ones, and %the opposite for $\tilde\Gamma^M$.
Further $\Lambda$ is the product $\Lambda=\Gamma^{890}$ of gamma matrices and $\varepsilon_{ABC}$ is the anti-symmetric symbol. The trace is taken over the colour indices which we have not written out. Further conventions of gamma matrices can be found in appendix A of \cite{Minahan:2015jta}.

The action above is invariant under the supersymmetry transformations 
\bea 
\delta_\eps  A_M &=& \eps \G_M \Psi \notag \, ,\nn\\
\delta_\eps \Psi &=& \frac{1}{2} F_{MN}\G^{MN}\eps + \frac{8}{7} \G^{\mu B} \phi_B \nabla_\mu \eps\, , \label{susytranson}
\eea
where $\mu =1,\dots, 7$. These transformations work as long as we have a 10-dimensional Majorana-Weyl spinor $\eps$ satisfying the generalised Killing spinor equation
\begin{equation} \label{killcond}
\nabla_\mu \eps  =  \frac{1}{2} \Gt_\mu \La \eps\, .
%\nabla_\mu \eps  =  \frac{1}{2r} \Gt_\mu \La \eps\, .
\end{equation}
 
For $S^7$ such an $\eps$ can be constructed using the conformal Killing spinors of the sphere \cite{Minahan:2015jta}, and in \cite{Polydorou2017} it was shown that  $\eps$ can be constructed for any compact 7D manifold admitting a pair of positive Killing spinors \eqref{Killing_spinor_7D}. Manifolds admitting positive Killing spinors have been classified in the mathematics literature \cite{Bar}. In 7D they fall into the categories listed in the introduction. 

The supersymmetry can be taken off-shell and for the case of Sasaki-Einstein (SE) manifolds (including the 3-Sasakian manifolds and of course $S^7$). A cohomological complex was found in \cite{Minahan:2015jta, Polydorou2017}. We provide the minimal amount of geometrical intuition of the complex. The SE manifolds are contact with a Reeb vector field $R$. Transverse to $R$, there is a K\"ahler (it is actually K\"ahler Einstein) structure. Using this transverse K\"ahler structure, one has a convenient representation of spinors $\Omega^{(0,\sbullet)}_H$, i.e. $(0,i)$-forms transverse to $R$ (as implied by the subscript $H$). The solutions to \eqref{Killing_spinor_7D} are the two parallel sections of $K^{\pm1/2}$ where $K$ is the line bundle $\Go_H^{(0,3)}$. One should compare 7D with the 5D case described in \cite{Qiu:2013pta} where the solutions to \eqref{Killing_spinor_7D} would have $\pm i/2$. 
In this spin representation, the fermions decomposes into various sections of $\Go_H^{(0,\sbullet)}$ and their conjugate. The bosonic partners also fall into the same pattern. We remark that although supersymmetry can be taken off-shell also for proper $G_2$-manifolds (see e.g. \cite{Prins:2018hjc}), the cohomological complex would be of a different nature compared to the Sasaki-Einstein case above since there is only one solution to \eqref{Killing_spinor_7D}. We hope to return to the proper $G_2$ case in future work.  

\subsection{The one-loop}\label{sec:one-loop}
Localisation argument then says that we compute one-loop round an instanton background.  But as we still lack a good understanding of instantons in 7D, we study only the trivial background.
 Under such background, all fields break up into sections of the complex $\Go_H^{(0,\sbullet)}\otimes\FR{g}$ valued in the adjoint and their conjugates.
The one-loop computation then reduces to a super-determinant \footnote{In this paper we ignore the phase of the super determinant. In 3D  the careful computation of the phase 
 will account for the shift in Chern-Simons level, but in 7D we leave this important detail for future work.}
\begin{align} \label{partfun2}
Z^{pert}
&= \int\limits_{\FR{g}} d\sigma e^{-\frac{24}{g^2_{7D}} V_7 \Tr(\sigma^2)}  \Big|{\sdet}'_{\Omega_H^{(0,\bullet)}\otimes\FR{g}}(-\L_R + iG_{\sigma})\Big|\, .
\end{align}
The integral is taken over the Lie algebra $\FR{g}$, while $V_7$ denotes the volume of the 7D manifold. The Lie derivative $\L_R$ acts on $\Go^{(0,p)}_H\otimes\FR{g}$, as does the gauge transformation $iG_{\gs}$ for constant $\gs$. We take the super-determinant of $-\L_R+iG_{\gs}$ over $\Go^{(0,p)}_H\otimes\FR{g}$ and the prime means we exclude the zero modes.

Since the determinant of $iG_{\gs}$ is rather standard, the main task is then to compute the superdeterminant
\begin{equation}
\sdet_{\Omega_H^{(0,\bullet)}}(-\L_R + x) \nn \, ,
\end{equation}
for some arbitrary constant $x$.
The usual trickery in super-symmetry shows that the super-determinant over $\Omega_H^{(0,\bullet)}$ descends to the cohomology $H_{\bar\partial_H}^{(0,\bullet)}$ where $\bar\partial_H$ is the transverse Dolbeault operator. So we focus on 
\begin{equation}
\sdet_{H_{\bar\partial_H}^{(0,\bullet)}}(-\L_R + x) \label{eq:sdet} \, .
\end{equation}
 To come this far, we require only the SE structure. But to compute this (still infinite dimensional) super-determinant, we will need to use equivariant index theorem, which works best if we have toric SE structure or hyper-toric structure.

The detailed index calculation in the toric SE case in 5D is provided in \cite{exotic_instanton}, the 7D case is not so different. The index calculation in the 7D hyper-toric case will be detailed in a separate publication \cite{HTaSFI}, along with other geometrical features of the hyper-toric manifolds.  

 For this article we state without proof (though similar arguments can be found in \cite{Schmude:2014lfa}) that 
$H_{\bar\partial_H}^{(0,0)}$ can be identified with holomorphic functions on the cone $C(X)$ over the 7D manifold $X$. Furthermore $(H_{\bar\partial_H}^{(0,3)})^*$ is isomorphic to $H_{\bar\partial_H}^{(0,0)}$ with the isomorphism provided by multiplication by the holomorphic volume form\footnote{we recall that
when $X$ is toric SE then $C(X)$ is toric Calabi-Yau, while if $X$ is hyper-toric 3-Sasakian then $C(X)$ is hyper-toric hyper-K\"ahler, see sec.\ref{sec:hypertoric3S}. In either cases, there is a holomorphic volume form that respects all the torus actions.}. Finally $H_{\bar\partial_H}^{(0,1)}=H_{\bar\partial_H}^{(0,2)}=0$.
%With this understanding, computing $\sdet(\L_R)$
%The contribution of 

For the case of toric SE manifolds, discussed e.g. in \cite{Polydorou2017}, these holomorphic functions are in one-to-one correspondence with the integer lattice points inside a certain cone determined by the toric data. The Lie derivative $\L_R$ acts on each holomorphic function with a weight that can be read off from the lattice point. Thus the super-determinant over $H_{\bar\partial_H}^{(0,0)}$ can be written as a regulated infinite product of weights of each lattice point in the cone.
 As for $H_{\bar\partial_H}^{(0,3)}$, the isomorphism mentioned above says that its weights under $\L_R$ is negative that of $H_{\bar\partial_H}^{(0,0)}$ (due to the complex conjugation) and shifted by the weight of the holomorphic volume form. This results in an infinite product of weights of lattice points over the interior of the negative cone.
Such a description of the super-determinant led to the definition of a generalised quadruple sine function $S_4^{C_\mu(X)}$ where $C_\mu(X)$ is the moment map cone determined by the torus action, see \cite{Winding:2016wpw}. Recall that the standard quadruple sine function \cite{Narukawa} is associated to the cone that is the first orthant of $\BB{R}^4$, which incidentally is also the moment map cone ${C_\mu(S^7)}$ of the seven sphere. So we see a nice generalisation from $S^7$ to the other toric Sasaki-Einstein manifolds.

To summarise, the perturbative part of the partition function for supersymmetric Yang-Mills on a 7D toric Sasaki-Einstein manifold $X$ is then \cite{Polydorou2017}
\begin{align}
Z^{\text{pert}}
&= \int\limits_t d \sigma \,  e^{-\frac{24}{g^2_7} V_7 \Tr(\sigma^2)} \prod_{\beta \neq 0} \Big|S_4^{C_\mu(X)}( i \bra \sigma, \beta \ket |\vec{R})\Big|\, , \label{pertparfun}
\end{align}
where $i\bra\gs,\gb\ket$ replaces $x$ above with $\beta$ being the non-zero roots of the Lie algebra $\FR{g}$ with Cartan subalgebra $\FR{t}$, and $\vec R$ provides the weights of $\L_R$ for each lattice point.

In this paper we will derive an analogous result for 7D \emph{hypertoric} 3-Sasakian manifolds. While these manifolds are also Sasaki-Einstein, they are not necessarily toric and so different techniques have to be used. This paper generalises the `proof-of-concept' calculation in \cite{Rocen2018} to arbitrary 7D hypertoric 3-Sasakian manifolds, which we now describe.

\section{Hypertoric 3-Sasakian manifolds} \label{sec:hypertoric3S}

In this section we briefly review 3-Sasakian and hypertoric geometry. We refer the reader to \cite{Boyer1998, Bielawski2000, Proudfoot2008survey} for comprehensive introductions to these topics.

A manifold $X$ is 3-Sasakian if its metric cone $C(X)$ is hyper-K\"ahler (HK).
Recall that the metric cone is defined as $C(X) = X \times \R^+$, with metric $ds^2_{C(X)} = dr^2 + r^2 ds^2_X$, where $r$ is the $\R^+$-coordinate and $ds^2_X$ the metric on $X$.
We will often view $X$ as the hypersurface in $C(X)$ where $r=1$.

Being hyper-K\"ahler means that $C(X)$ has three complex structures $I,J,K$ satisfying the quaternionic relations
\begin{equation}
I^2=J^2=K^2=IJK=-1 \, ,
\end{equation}
and the metric is K\"ahler with respect to each of these. We thus also have three symplectic forms.
From these structures on $C(X)$ we obtain three Reeb vector fields and contact forms on $X$ via
\begin{align}
R_a &= I_a(r \partial_r)|_{r=1}, \quad I_a=I,J,K \,,\\
\kappa_a(Y) &= g(R_a,Y)\, .
\end{align}
These form a 3-Sasakian structure on $X$ and satisfy the relations
\begin{align}
\iota_{R_a} \kappa_b &= \delta_{ab} \, , \\
[R_a,R_b] &= \epsilon_{abc}R_c \, .
\end{align}

An HK manifold of dimension $4n$ is said to be \emph{hypertoric} if it admits an effective action of the torus ${\mathbb T}^n$ that is Hamiltonian with respect to each of the three symplectic structures. We call a 3-Sasakian manifold of dimension $4n-1$ \emph{hypertoric} if its HK cone is hypertoric. Equivalently, if it admits an effective action of ${\mathbb T}^n$ that preserves the 3-Sasakian structure \cite{Boyer1998}.\footnote{Note that in \cite{Boyer1998} such manifolds are called `toric' 3-Sasakian manifolds.}

Boyer and Galicki \cite{Boyer1998} constructed such hypertoric 3-Sasakian manifolds via a 3-Sasakian quotient, i.e. by taking toral reductions of 3-Sasakian spheres. We prefer to view the hypertoric 3-Sasakian manifolds as hypersurfaces in their HK cones and construct the latter by HK quotients as in \cite{Bielawski2000}. These constructions are of course very much related. % and we can borrow many results from \cite{Boyer1998}.

%\ybox{Insert somewhere: By a theorem of Bielawski (insert ref) all hypertoric 3-Sasakian manifolds can be constructed via such quotients. }

Let $\H^n$ be the quaternionic vector space with the standard hyper-K\"ahler structure given by the \emph{right} multiplication by $i,j,k$. We think of $\H^n$ as $\C^n \times (\C^n)^*$ with coordinates $\vec{q}= \zv + j \wv$. The torus ${\mathbb T}^n$ acts as \emph{left} multiplication on $\H^n$ via $\vec q\mapsto t\vec q$ or $(\zv, \wv)\mapsto(t\zv, t^{-1}\wv)$. This action preserves by design the HK structure.
Let ${\mathbb K}$ be a subtorus of ${\mathbb T}^n$ of rank $k$ with its Lie algebra denoted $\mathfrak{k}$. We have the following exact sequences:  
\begin{equation}
\begin{tikzcd}[column sep=small]
0 \arrow{r} &  \mathfrak{k} \arrow{r}{\iota} & \mathfrak{t}^n \arrow{r}{\beta} & \mathfrak{t}^d \arrow{r} & 0 
\end{tikzcd}  \, ,\label{eq:exactseq1}
\end{equation}
\begin{equation}
\begin{tikzcd}[column sep=small]
0 \arrow{r} &  (\mathfrak{t}^d)^* \arrow{r}{\beta^*} & (\mathfrak{t}^n)^* \arrow{r}{\iota^*} & \mathfrak{k}^* \arrow{r} & 0 
\end{tikzcd} \, , \label{eq:exactseq2}
\end{equation}
where $\mathfrak{t}^d$ is the Lie algebra of ${\mathbb T}^d={\mathbb T}^n/{\mathbb K}$.
We represent the map $\iota^*$ by an integer $n \times k$ weight matrix $Q$. The moment maps for the subtorus action are 
\begin{align}
\mu^a_\R(\vec{z},\vec{w}) &= -\frac12 \sum\limits_{i=1}^n \left( |z_i|^2-|w_i|^2 \right) Q_i^{a}  + c_1^a \, ,\label{mu_3} \\
\mu^a_\C(\vec{z},\vec{w}) &= i \sum\limits_{i=1}^n z_iw_i Q_i^{a}  +c_2^a + ic_3^a\, ,\label{mu_complex}
\end{align}
where $a=1,\dots, k$. The $c$'s are in principle arbitrary for now, but soon we will set them to zero since we aim eventually for a cone structure for the HK reduction, which requires invariance under simultaneous scaling over all $z,w$.
Taking the HK quotient $\mu^{-1}(0)/{\mathbb K}$ gives a hypertoric variety of dimension $4d=4(n-k)$ admitting a hypertoric action of ${\mathbb T}^d$.

\subsection{Hyper-plane arrangement}\label{sec:Hpa}
Hypertoric varieties can be described by hyperplane arrangements \cite{Bielawski2000}, which we now explain. The map $\beta$ in \eqref{eq:exactseq1} can be described by an integer $n \times d$ matrix $\Qt$, whose rows can be thought of as $n$ vectors $\{v_i\}$ in $\R^d$. Moreover, we can write the central elements as $c_k^a = \sum_{i=1}^n \lambda_k^i Q_i^a$, $k=1,2,3$,  to obtain $n$ triplets of scalars.
We then define $n$ triplets of hyperplanes in $\R^d$ via
\begin{equation}
H_k^i = \{ y\in \R^d | y \cdot v_i = \lambda_k^i \} \,.
\end{equation}
Combining each triplet we obtain $n$ hyperplanes $H^i = H^i_1 \times H^i_2 \times H^i_3$ in $\R^{3d}$. 

Many geometrical properties of hypertoric varieties can be expressed in terms of properties of these hyperplanes, similar in spirit to how properties of toric varieties can be stated in terms of fans.

We are interested in the special case of HK varieties that are cones over compact smooth 3-Sasakian manifolds. From \cite{Bielawski2000} we obtain such manifolds by setting all $\lambda_k^i=0$ and the smoothness implies the following properties for the normal vectors $\{v_i\}$:
\begin{enumerate}
 \item any collection of $d$ vectors from $\{v_i\}$ are linearly independent and \item  any collection of less than $d$ vectors from $\{v_i\}$ can be completed to a $\Z$-basis of $\Z^d$.
 \end{enumerate}
If these conditions are satisfied, we can then view the 3-Sasakian manifold as the hypersurface where
\begin{equation}
 \sum_{i=1}^n \left( |z_i|^2+|w_i|^2 \right) = 1 \, .\label{|mu|}
\end{equation}

In this paper we are interested in 8D cones, i.e. the case when $d=2$. In this case the conditions simplifies: 
\begin{enumerate}
\item the vectors $\{v_i\}$ are pairwise linearly independent
\item each $v_i$ has relatively prime components 
\end{enumerate} 
Since the $v$'s are the rows of the matrix $\Qt$ we may equally well state this as a property of the matrix $\Qt$. 

We may also state such conditions in terms of the weight matrix $Q$. Firstly, note that many weight matrices define the same subtorus action. We are free to choose any basis of $\mathfrak{k}$, meaning that any two weight matrices related by a $GL(k,\Z)$ transformation give the same quotient. Furthermore, we may permute the rows of $Q$ (i.e. permute the index $i$) or flip the signs of any row (i.e. $z_i\to w_i$, $w_i\to -z_i$). By reparametrising the one-parameter subgroups we may also take the greatest common divisor of the elements in each column to be 1.

We recall the following notions from \cite[sec.13.7.2]{Boyer1998}\footnote{Note that the matrix we call $Q$ is the transpose of the matrix called $\Omega$ in \cite{Boyer1998}.}: $Q$ is called \emph{non-degenerate} if all of its $k \times k$ minors have non-zero determinants. For a non-degenerate matrix $Q$, let $g$ be the gcd of all the $k \times k$ minor determinants. By reparametrising $Q$ as discussed above we may assume that $g=1$.
A non-degenerate matrix $Q$ is called \emph{admissible} if each of its $(k+1) \times k$ minors can be completed into an $SL(k+1,\Z)$-matrix.\footnote{A $(k+1) \times k$ matrix can be completed into an $SL(k+1,\Z)$-matrix iff the determinants of its $k\times k$ minors have gcd $1$.}
If $Q$ is non-degenerate and admissible, then it gives rise to a free action and $C(X) = \mu^{-1}(0)/{\mathbb K}$ is the cone over a smooth 3-Sasakian manifold $X$.
%
%\rbox{I think this is the statement... But need to show that it is equivalent to Bielawski-Dancers statement... \\
%Actually, this statement in BoyerGalicki is about if you FIRST restrict to $ \sum_{i=1}^n \left( |z_i|^2+|w_i|^2 \right) = 1 $, THEN do the quotient. need to check the order of this doesn't matter. Probably just follows  by interpreting restriction as moment map of right action by i.}
%
%\ybox{Also mention that non-degeneracy implies that at most n-k-1 quaternionic coordinates can vanish simultaneously on the zero set of the moment map? but maybe this is only important for the index calculation?}

The data $\{v_i\}$ and $Q$ are equivalent. In fact, from $Q$ regarded as an $n\times k$ matrix, complete it into an $SL(n,\BB{Z})$ matrix and invert it. Then the first $d=n-k$ rows of the inverse is the matrix $[v_1,\cdots,v_n]$. The $v$'s determined this way are independent of the completion of $Q$ up to an overall $SL(d,\BB{Z})$ action. 
Conversely one takes the matrix $[v_1,\cdots,v_n]$, after completing it and inverting it, one recovers $Q$ as the last $k$ columns of the inverse. 

In the following, we will focus more on the hyper-plane arrangement point of view, i.e. the $v_i$'s are more important to us.

\subsection{Geometry close to a torus fixed locus}\label{sec_Gctatfp}

We recall from sec.\ref{sec:hypertoric3S} that we have a left ${\mathbb T}^d$ ($d=2$ now) action on $X$. There is also a natural right action by $SU(2)$ on $X$ as follows. Starting from $\BB{H}^n$, one can multiply all the $\vec q\in\BB{H}^n$ with a unit-quaternion (see the notation of sec.\ref{sec:hypertoric3S}). This action preserves the moment map conditions \eqref{mu_complex} \eqref{mu_3} for $c_{1,2,3}=0$.
It clearly commutes with the left $U(1)$ actions and so descends to the HK cone. Finally the action preserves \eqref{|mu|} and so descends to the 3-Sasakian manifold $X$.

The $SU(2)$ action is locally free but may have finite subgroups $\Gc\subset SU(2)$ as stability groups. Generically $\Gc$ is trivial or $\BB{Z}_2$ (generated by $\pm1\in SU(2)$). The latter case can happen if
\bea \sum_{i=1}^nv_i={\rm even~ vector}.\nn\eea
We give a quick explanation. The right multiplication with $-1\in SU(2)$ on $\BB{H}^n$ can sometimes be undone by a left action of ${\mathbb K}\subset {\mathbb T}^n$. This 
can happen iff $[-1,\cdots,-1]$ is contained in ${\mathbb K}$. This condition is equivalent to the one above expressed entirely in terms of the data of hyper-arrangements i.e. the $v$'s). 
We do point out that even when the fibre is $SO(3)$ the manifold $X$ is still simply connected. This follows from a rather involved computation of $\pi_1$ done in sec.13.7.6 in \cite{BoyerGalicki}.

\smallskip

So far we have dealt with a generic fibre, now we gather more information about the geometry close to a degenerate fibre, i.e. close to the ${\mathbb T}^2$ fixed locus. 
These loci are where one of the $q$ in $\BB{H}^n$ vanishes. We start with an example. 

Take $n=3$ and ${\mathbb K}\subset {\mathbb T}^3$ acts with weight matrix $Q=[1,3,2]$. At the locus $q_1=0$, solving the moment map condition gives 
\bea [q_2,q_3]=[(2/5)^{1/2}e^{i\ga}\hat q,(3/5)^{1/2}e^{i\gb}j\hat q],~~\hat q\in \BB{H},~|\hat q|=1.\nn\eea
This shows that this locus is the right-$SU(2)$ orbit of the point $[(2/5)^{1/2}e^{i\ga},(3/5)^{1/2}e^{i\gb}j]$. But the $SU(2)$ action is not free, e.g. the right multiplication by $e^{2i\pi/5}$ can be undone by a left action of $\BB{K}$: $[q_2,q_3]\to [e^{-12i\pi/5}q_2, e^{-8i\pi/5}q_3]$. In fact the element $e^{2i\pi/5}$ generates the entire stability group $\BB{Z}_5$. 
%The fibre at this locus is thus $L(5,1)=S^3/\BB{Z}_5$, where $L(p,q)$ denotes the lens space obtained from $S^3$ by the action $(z_1,z_2)\mapsto (e^{2\pi i/p}z_1,e^{2\pi iq/p}z_2)$.

To understand the local geometry better, we find the ${\mathbb K}$-invariant coordinates. 
We choose them differently according to whether $\hat q$ is in the neighbourhood of $e^{i\gt}$ or it is close to $je^{i\gt}$ (in fact if one mods out $e^{i\gt}$ from the right, the two loci correspond to the north and south pole of the resulting $\BB{C}P^1=S^3/U(1)$) 
\bea {\rm north}:&& u_n=w_2/\bar z_2,~~~\tau_n=e^{i\arg z_2^2w_3^3},~~~a_n=(1+|u_n|^2)^{1/2}\nn\\
&& \xi_{n1}=a_n^{-1}(z_1+\bar w_1u_n)e^{-i\arg(z_2w_3)},~\xi_{n2}=a_n^{-1}(w_1-\bar z_1u_n)e^{i\arg(z_2w_3)},\label{adp_wght_ex_n}\\
{\rm south}:&& ~u_s=\bar z_2/w_2,~~~\tau_s=e^{i\arg w_2^{-2}z_3^{-3}},~~~a_s=(1+|u_s|^2)^{1/2}\nn\\
&&\xi_{s1}=a_s^{-1}(z_1u_s+\bar w_1)e^{i\arg(w_2z_3)},~~~\xi_{s2}=a_s^{-1}(w_1u_s-\bar z_1)e^{-i\arg(w_2z_3)}.\label{adp_wght_ex_s}\eea 
Going from north to south we have the transition function
\bea & u_s=u_n^{-1},~~~\tau_s=-\tau_ne^{-5i\arg u_n},\nn\\
&\xi_{s1}=-\xi_{n1}e^{i\arg u_n},~~~\xi_{s2}=-\xi_{n2}e^{-3i\arg u_n}.\nn\eea
We see from the transition that $(u,\tau)_{n,s}$ parametrises the total space of
a circle bundle over $\BB{C}P^1$ of degree 5. With $u_{n,s}$ being the standard complex coordinates of $\BB{C}P^1$ and $\tau_{n,s}$ is the $S^1$ fibre coordinate.
We introduce the following notation 
\bea 
\begin{tikzpicture}
  \matrix (m) [matrix of math nodes, row sep=1em, column sep=1.5em]
    {  {\cal S}(n)   \\
         \BB{C}P^1  \\ };
  \path[->]
  (m-1-1) edge node[left] {\scriptsize{$\pi$}} (m-2-1);
\end{tikzpicture}\nn\eea
for the circle bundle over $\BB{C}P^1$ of degree $n$, in particular $S^3={\cal S}(-1)$. We write ${\cal S}(n)_{\BB{C}}$ for the associated $\BB{C}^*$ bundle, in particular $\BB{C}^2\backslash\{0\}={\cal S}(-1)_{\BB{C}}$.
As for the holomorphic line bundle over $\BB{C}P^1$ with degree $n$, we stick to the standard notation ${\cal O}(n)$. We do however abuse the notation by writing the line bundle associated with ${\cal S}(n)$ also as ${\cal O}(n)$, since these two concepts coincide on $\BB{C}P^1$, both of which uniquely fixed by the first Chern class.

For later use, we write ${\cal S}(n)$ as an associated bundle: since $U(1)$ can act from the left and right on $S^3$, we take extra caution to specify that our ${\cal S}(n)$ is
\bea {\cal S}(n)= S^1\times_{U(1)}S^3,\nn\eea 
where $U(1)$ acts on $S^3$ by left multiplication and on $S^1$ with weight 5. 
The global right $SU(2)$ action then acts on ${\cal S}(n)$ as right multiplication on $S^3$.

The coordinates $\xi_{1,2}$ parametrise the infinitesimal deformation away from $q_1=0$. From the transition function, we see that the normal bundle splits into two line bundles
\bea N\simeq {\cal O}(1)\oplus {\cal O}(-3),\nn\eea
Note $\xi_{1,2}$ are the fibre coordinates so they would transform in the corresponding dual line bundle.
From the detailed expression of $\xi_{1,2}$, one also sees the twistor space structure: e.g. as $u_n$ goes from $0$ to $\infty$ the $\xi$-coordinate goes from $z_1$ to $\bar w_1$, showing the familiar picture that the complex structure transverse to the fibre varies across $aI+bJ+cK$ for $a^2+b^2+c^2=1$ as one traverses the fibre.

In our way of presenting the local geometry, that is, in making the right $SU(2)$ action explicit, we necessarily break holomorphicity. This is because in sec.\ref{sec:hypertoric3S} we have chosen to work with the complex structure induced by the right multiplication by $i$. This choice will be (by design) rotated by the right $SU(2)$ action. 

Seeing that we are mainly interested in enumerating the holomorphic functions on the HK cone, it is more useful to present the local geometry keeping holomorphicity. To this end, one needs to slightly modify the HK quotient in sec.\ref{sec:hypertoric3S}. The moment map condition \eqref{mu_complex} is holomorphic but \eqref{mu_3} is not. So instead of enforcing \eqref{mu_3} and mod out by the subtorus ${\mathbb K}$, we remove the points that would be excluded by the moment map conditions (the so called unstable points) and mod by ${\mathbb K}_{\BB{C}}$, the complexification of ${\mathbb K}$. Also we do not impose \eqref{|mu|} to access the entire HK cone. This way we keep the holomorphicity explicit. 

At the torus fixed locus $q_1=0$, we have the local geometry
\bea \BB{C}^*\times_{\BB{C}^*}(\BB{C}^2\backslash\{0\}),\label{eq:used}\eea
where $\BB{C}^2\backslash\{0\}\simeq {\cal S}(-1)_{\BB{C}}$ is parametrised as $z+jw$. It is thought of as the cone over $S^3$, i.e. the simultaneous scaling of $z,w$ (right multiplication by $\BB{C}^*$) is the cone direction, it is also the cone direction of the HK cone. 
By contrast, in the notation $\times_{\BB{C}^*}$, the $\BB{C}^*$ acts on $z+jw$ as left-multiplication while on the left $\BB{C}^*$ factor with weight 5. The normal bundle to the locus $q_1=0$ has a similar presentation
\bea \BB{C}^*\times_{\BB{C}^*}({\cal S}(-1)_{\BB{C}}\oplus{\cal O}(1)\oplus{\cal O}(1))\label{local_picture_ex}\eea
where ${\cal S}(-1)_{\BB{C}}\oplus{\cal O}(1)\oplus{\cal O}(1)$ is the fibre product of three bundles (we follow the standard notation in \cite{husemoller1994fibre} and use $\oplus$ for fibre product, even though ${\cal S}(-1)_{\BB{C}}$ is not a vector bundle).
The $\BB{C}^*$ acts on ${\cal S}(-1)_{\BB{C}}$ as it does before, it also acts on the fibre of the latter two ${\cal O}(1)$ with weights $-1,3$.

We list also the holomorphic functions 
\bea {\rm north}:&& \tau_n=z_2^2w_3^3,~~u_n=w_2z_2,~~\xi_{n1}=z_1/(z_2w_3),~~\xi_{n2}=w_1z_2w_3.\label{wght_ex_n}\\
{\rm south}:&& \tau_s=w_2^{-2}z_3^{-3},~~u_s=w_2z_2,~~\xi_{s1}=z_1z_3w_2,~~\xi_{s2}=w_1/(w_2z_3)\label{wght_ex_s}.\eea 
We intentionally used the same notation as in \eqref{adp_wght_ex_n} and \eqref{adp_wght_ex_s}. This is not only because the functions with the same symbol in both tables have the same $U(1)$ weights, there is another more speculative reason that will be used in sec.\ref{sec:Agiaas}.

%
%Going from northern to southern hemisphere 
%\bea & u_s=u_n^{-1},~~~\tau_s=-\tau_ne^{-5i\arg u_n},\nn\\
%&\xi_{s1}=-\xi_{n1}e^{i\arg u_n},~~~\xi_{s2}=-\xi_{n2}e^{-3i\arg u_n}\nn\eea
%plus another coordinate $u=z_2w_2$ which is common for north and south.
%Now $\tau$ is the fibre coordinate of a holomorphic $\BB{C}^*$ bundle while $u,z,w$ are the fibre coordinate of holomorphic line bundles. The degree of these bundles are $5,\;-1,\;3,\;2$ respectively.

The way we read off the local geometry by way of finding the local invariant coordinates may seem a bit cumbersome, but the procedure can be done entirely in terms of the hyper-plane arrangement. We give a table for the weights for general case (again omitting details of the derivation).
For each $v_i$, we define a very important vector $P_i$ as a linear combination of all the $v$'s with coefficient $\pm 1$ as follows.
Draw a directed line along $v_i$. All vectors $v_j$ to the left of this line get a plus sign, and all the $v_j$'s to the right get a minus sign. 
In formula we can write this as
\begin{equation}
P_i = v_i + \sum_{j \neq i} \sgn (\det [v_i,v_j]) v_j  \,. \label{eq:Rdef}
\end{equation}
See figure \ref{fig:normalvectorsR}, we shall have more use of $P_i$ in sec.\ref{sec:Hfitolpiac} as a bookkeeping device for the holomorphic functions. 
We also pick for each $v_i$ any integer vector $\hat v^i$ such that $\hat v^i\cdotp v_i=1$, we let $q_i=\hat v^i\cdotp P_i$, $r_i=v_i\times P_i$ (so $q_i$ is only well-defined mod $r_i$).

There is one ${\mathbb T}^2$ fixed locus for each $i$, which has geometry the 
same as in the picture \eqref{local_picture_ex}, i.e.
\bea \BB{C}^*\times_{\BB{C}^*}({\cal S}(-1)_{\BB{C}}\oplus{\cal O}(1)\oplus{\cal O}(1))\label{local_picture}\eea
where in writing $\times_{\BB{C}^*}$, the $\BB{C}^*$ acts on the leftmost $\BB{C}^*$ with weight $r_i$. It still acts on ${\cal S}(-1)_{\BB{C}}$
in the same way, but with weights $q_i$ and $2-q_i$ on the fibre of ${\cal O}(1)\oplus{\cal O}(1)$. 

From these quantities we make a table of the weights of the holomorphic functions.
%We pick $\vec\go=[\go_1,\go_2]$ to be the equivariant parameter   
\bea \begin{array}{|c|c|c|}
\hline
& \textrm{north} & \textrm{south} \\
\hline 
\xi_1 & [\hat v^i,q_i] & [\hat v^i,2-q_i]\\
\hline
\xi_2 & [-\hat v^i,2-q_i] & [-\hat v^i,q_i]\\
\hline
\tau & [Jv_i,r_i] & [Jv_i-r_i] \\ 
\hline u & [\vec 0,2] & [\vec 0,2] \\ 
\hline
\end{array}\label{tbl_hol_fun}\eea
where $J$ is the standard complex structure on $\BB{R}^2$ (counter clockwise rotation by $\pi/2$). To read the table, the first two entries are the weights of each function under the residue $\BB{T}^2$ action from ${\mathbb T}^n$ after $\BB{K}={\mathbb T}^{n-2}$ has been mod out. We pick explicit basis of weights of this $\BB{T}^2$ and call them $e_1$ and $e_2$-weight. 
The last entry is the weight under the right multiplication by $\BB{C}^*$. We call this weight the R-weight (as it is actually the R-symmetry).

Note also that the weights of $\xi_{1,2}$ are well-defined up to shifts by weights of $\tau$, since there is freedom in adding to $\hat v^i$ multiples of $Jv_i$. 
Our concrete example earlier corresponds to $v_1=[1;0]$, $v_2=[-1;2]$ and $v_3=[1;-3]$ and $P_1=[-1;5]$. One can easily recover the weights of functions in \eqref{wght_ex_n}, \eqref{wght_ex_s}.

Or if one prefers to make the $SU(2)$ action explicit and sacrifice holomorphicity, then the local geometry is
\bea {\cal O}(-q_i)\oplus {\cal O}(q_i-2)\oplus {\cal S}(r_i).\label{local_picture_adpt}\eea
All three bundles are associated to $S^3$ from the left while $SU(2)$ acts from the right. 

%Our concrete example earlier corresponds to $v_1=[1;0]$, $v_2=[-1;2]$ and %%$v_3=[1;-3]$ and $P_1=[-1;5]$.
%total space of 
%\bea {\cal O}(r_i) \to \BB{C}P^1\nn\eea
%with an $SU(2)$ action. To visualise the $SU(2)$ action more clearly, we write ${\cal O}(r_i)$ as an associated circle bundle over $S^2$
%\bea (S^1\times_{U(1)}S^3) \circlearrowright SU(2),\label{tp_fibre}\eea
%where the $U(1)$ acts on $S^1$ with weight $r_i$ and acts on $S^3$ as left multiplication. There is a natural right $SU(2)$ action on $S^3$ that commutes with the left $U(1)$ action and hence it acts on the total space of ${\cal O}(r_i)$. If we identity the tubular neighbourhood of this fibre with its normal bundle, then it splits into two line bundles 
%\bea {\rm normal~bundle}\simeq L_{-q_i}\oplus L_{q_i-2}\label{normal_bdl}\eea
%where $L_n$ is the line bundle associated with ${\cal O}(n)$.
This picture of the local geometry round a torus fixed fibre will have huge bearing on our factorisation results, as the geometry will dictate the constituents in our factorisation. 

\section{Enumerating holomorphic functions} \label{sec:counting}
It was recalled in sec.\ref{sec:one-loop} that the 1-loop part of the localisation involves enumerating the holomorphic functions on the hyper-K\"ahler cone and computing their weights under the Lie derivative $L_R$. We begin with a manual enumeration in sec.\ref{sec:De}. Then we will organise the result in terms of a cone in $\BB{R}^3$ by using solely the data of hyper-plane arrangement. 

\subsection{Direct enumeration}\label{sec:De}
In \cite[section~6.1]{Bullimore2016} it is described how the ring of holomorphic functions is determined from the matrix $\Qt$. We review this construction here.

The generators of the ring of holomorphic functions come in two types, those that are neutral under the residual ${\mathbb T}^d$ action and those that are charged. The neutral generators are of the form $N_i=z_i w_i$. The charged ones are obtained as follows: we write down a monomial $m_a=\prod_{i=1}^nz_i^{\Qt^i_a}$ for fixed index $a$. As $\Qt$ represents the map $\gb^*$ in the exact sequence \eqref{eq:exactseq2}, it is clear that $m_a$ is neutral under the sub-torus ${\mathbb K}$ as it should if $m_a$ were to be a function on the quotient. Since certain entries $\Qt_a^i$ can be negative so $m_a$ can be singular. But if certain $\Qt_a^i<0$ then  we replace $z_i^{\Qt_a^i}$ in $m_a$ with $w_i^{-\Qt_a^i}$ which keeps the neutrality but avoids the singularity at $z_i=0$.
This reasoning leads us to the definition:
for every element $A\in \Z^d$ of the ${\mathbb T}^d$ charge lattice, set
\begin{align}
C^A := \prod_{i=1}^n 
\begin{cases}
z_i^{|\tilde{Q}^i_A|}, & \text{if}\ \tilde{Q}^i_A>0 \\
w_i^{|\tilde{Q}^i_A|}, & \text{if}\ \tilde{Q}^i_A<0 
\end{cases} 
\label{eq:ringrelations}
\end{align}
where $\tilde{Q}_A := \tilde{Q} A \in \Z^n$.

The $C^A$ obey the multiplication relations
\begin{align}
C^A C^B = C^{A+B} \prod\limits_{i \text{ s.t. } \tilde{Q}_A^i\tilde{Q}_B^i < 0 } (z_iw_i)^{\text{min}(|\tilde{Q}^i_A|,|\tilde{Q}^i_B|)} \, .
\end{align}
This is also easy to understand. For example if $\tilde{Q}_A^i=-p$ and $\tilde{Q}_B^i=q$ with $q>p>0$, then $\Qt^i_{A+B}>0$. Thus $C^A$ would contain $w^p_i$, $C^B$ would contain $z_i^q$ while $C^{A+B}$ contains $z^{q-p}_i$. The mismatch between $C^AC^B$ and $C^{A+B}$ can be corrected with $(z_iw_i)^p$.

These $C^A$ and the $N_i= z_iw_i$ together generate the ring of holomorphic functions.
We grade them by their weights under the residual ${\mathbb T}^2$-action and the right multiplication by $U(1)$. Earlier, we have named these three weights as $e_1$, $e_2$ and $R$-weights respectively.

\begin{figure}[tbp]
\centering
\begin{tikzpicture}[scale=.7]
\draw [->] (-5,0) -- (5,0) node [below] {\small $s$};
\draw [->] (0,-5) -- (0,5) node [left] {\small $t$};
\draw [line width=0.25ex] (-4,0) -- (4,0);
\draw [line width=0.25ex] (0,-4) -- (0,4);
\draw [line width=0.25ex] (-2*1.5,3*1.5) -- (2*1.5,-3*1.5);
\foreach \x in {-4,...,4}
\foreach \y in {-4,...,4}
	{
	\draw [black,fill=black] (\x,\y) circle (0.5ex);
	 }
\node at (3.5,3.5) {\large I};
\node at (3.5,-1.5) {\large II};
\node at (0.5,-3.5) {\large III};
\node at (-3.5,-3.5) {\large IV};
\node at (-3.5,1.5) {\large V};
\node at (-0.5,3.5) {\large VI};
\end{tikzpicture}
\caption{Hyperplanes for the example discussed in the text superimposed on the ${\mathbb T}^2$ charge lattice. The six regions formed correspond to the six ways of resolving the absolute value signs in \eqref{eq:muweight}.}
\label{fig:hyperplaneregions}
\end{figure}
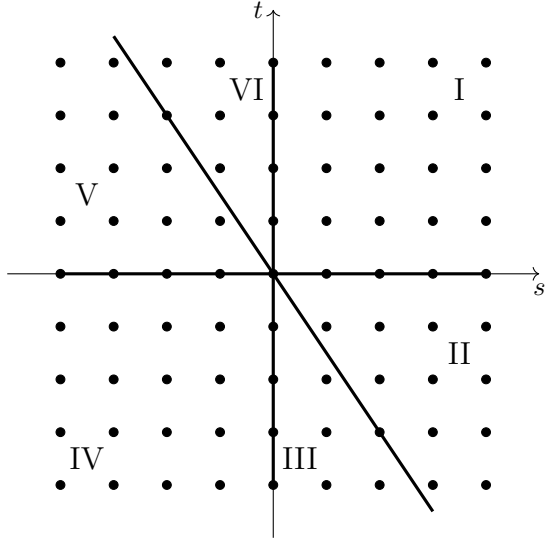

Let us turn to an example to illustrate this.
Let 
%\begin{equation}
%\Qt = \begin{pmatrix}
%1 & 0 & 3 \\
%0 & 1 & 2
%\end{pmatrix}
%\end{equation}
%
\begin{equation}
\Qt = \begin{pmatrix}
1 & 0 \\
0 & 1 \\
3 & 2
\end{pmatrix} \label{eq:Qtmatrix321} \,.
\end{equation}
The hyperplanes for this example are illustrated in figure \ref{fig:hyperplaneregions}.

%\begin{figure}[tbp]
%\centering
%\begin{tikzpicture}[scale=.5]
%\draw [->] (-5,0) -- (5,0) node [below] {\small $x$};
%\draw [->] (0,-5) -- (0,5) node [left] {\small $y$};
%\draw [line width=0.5ex] (-4,0) -- (4,0);
%\draw [line width=0.5ex] (0,-4) -- (0,4);
%\draw [line width=0.5ex] (-2,3) -- (2,-3);
%\end{tikzpicture}
%\caption{Hyperplane arrangement for the example discussed in the text.}
%\label{fig:hyperplanes}
%\end{figure}

Given a point $A=(s,t)$ in the ${\mathbb Z}^2$ lattice, we would get the following holomorphic functions:

\begin{align}
C^A N_1^m N_2^n  = (z_1 \text{ or } w_1)^{|s|} (z_2 \text{ or } w_2)^{|t|} (z_3 \text{ or } w_3)^{|3s + 2t|} (z_1 w_1)^m (z_2 w_2)^n \label{eq:holofuns} \, ,
\end{align}
where we pick $z_i$ or $w_i$ depending on if the exponent is positive or negative (before taking the absolute value). Here $m,n=0,1,\dots$. 
Let $e_1$ and $e_2$ be the U(1)'s of the ${\mathbb T}^2$-action and let $R$ be the U(1) of the Reeb. The functions in \eqref{eq:holofuns} have the following weights:
\begin{align}
&e_1\text{-weight}: \quad s \, ,\\
&e_2\text{-weight}: \quad t  \, ,\\
&R\text{-weight}: \quad |s|+|t|+|3s+2t|+2(m+n) \, .
\end{align}
We grade the holomorphic functions by their $R$-weight (degree), but also by their $e_1$ and $e_2$-weights.
For each $R$-weight $u$ we must ask what $s,t,m,n\in \Z$, $m,n\geq 0$ satisfy
\begin{equation}
u = |s|+|t|+|3s+2t|+2(m+n) . \label{eq:muweight}
\end{equation}
In order to determine this we need to resolve the absolute value signs. Where these absolute values switch sign is determined by the hyperplanes. e.g. to one side of the hyperplane with normal vector $(3,2)$ we have that $3s+2t$ is negative, and on the other side it is positive. We superimpose the hyperplane arrangement onto the $(s,t)$-lattice and each region enclosed by the planes corresponds to a specific way of resolving all the absolute values in \eqref{eq:muweight}, see figure \ref{fig:hyperplaneregions}.

\begin{figure}[tbp]
\centering
\begin{tikzpicture}[scale=.7]
\draw [->] (-5,0) -- (5,0) node [below] {\small $s$};
\draw [->] (0,-5) -- (0,5) node [left] {\small $t$};
\draw [line width=0.25ex] (-4,0) -- (4,0);
\draw [line width=0.25ex] (0,-4) -- (0,4);
\draw [line width=0.25ex] (-2*1.5,3*1.5) -- (2*1.5,-3*1.5);
\foreach \x in {-4,...,4}
\foreach \y in {-4,...,4}
	{
	\draw [black,fill=black] (\x,\y) circle (0.5ex);
	 }
\foreach \k in {7,5,3,1}
	{\draw [line width=0.25ex, color=black] (0,\k/3) -- (\k/4,0) -- (2/5*\k,-3/5*\k) --(0,-\k/3) -- (-\k/4,0) -- (-2/5*\k,3/5*\k) -- cycle;
	}
\node at (3.5,3.5) {\large I};
\node at (3.5,-1.5) {\large II};
\node at (0.5,-3.5) {\large III};
\node at (-3.5,-3.5) {\large IV};
\node at (-3.5,1.5) {\large V};
\node at (-0.5,3.5) {\large VI};
\node[circle,scale=0.65,color=white, fill=black] at (-2,3) {$\boldsymbol{2}$};
\node[circle,scale=0.65,color=white, fill=black] at (-1,3){$\boldsymbol{1}$};
\node[circle,scale=0.65,color=white, fill=black] at (-2,1){$\boldsymbol{1}$};
\node[circle,scale=0.65,color=white, fill=black] at (-1,1) {$\boldsymbol{3}$};
\node[circle,scale=0.65,color=white, fill=black] at (0,1) {$\boldsymbol{3}$};
\node[circle,scale=0.65,color=white, fill=black] at (1,1){$\boldsymbol{1}$};
\node[circle,scale=0.65,color=white, fill=black] at (-1,-1){$\boldsymbol{1}$};
\node[circle,scale=0.65,color=white, fill=black] at (0,-1) {$\boldsymbol{3}$};
\node[circle,scale=0.65,color=white, fill=black] at (1,-1) {$\boldsymbol{3}$};
\node[circle,scale=0.65,color=white, fill=black] at (2,-1){$\boldsymbol{1}$};
\node[circle,scale=0.65,color=white, fill=black] at (1,-3){$\boldsymbol{1}$};
\node[circle,scale=0.65,color=white, fill=black] at (2,-3) {$\boldsymbol{2}$};
\end{tikzpicture}
\caption{Holomorphic functions correspond to intersections between the polygons and the $\Z^2$-lattice. The multiplicities tell how many independent functions correspond to each point.}
\label{fig:PolygonPointsEx}
\end{figure}
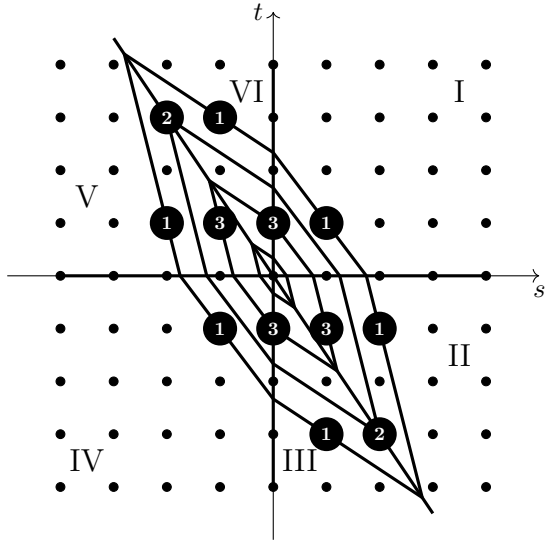

In each region we can now solve \eqref{eq:muweight}. In the example above, for Region I we look for solutions to
\begin{align}
u  = 4s+3t + 2(m+n) \,.
\end{align}
Let us say we are interested in $u=7$. In Region I we thus look at all integer lattice points $(s,t)$ satisfying
\begin{align}
4s+3t &= 7,  \quad \text{ with multiplicity 1, } (m+n=0), \\
4s+3t &= 5,  \quad \text{ with multiplicity 2, } (m+n=1), \\
4s+3t &= 3,  \quad \text{ with multiplicity 3, } (m+n=2), \\
4s+3t &= 1,  \quad \text{ with multiplicity 4, } (m+n=3).
\end{align}
The next equation, $4s+3t = -1$, has no solution in the region.
Geometrically, this corresponds to integer lattice points lying on the lines plotted in Region I in figure \ref{fig:PolygonPointsEx}. The multiplicities start at $1$ for the outermost line, is $2$ for the second outermost line, etc.
Repeating this procedure region-by-region we get the picture in figure \ref{fig:PolygonPointsEx}.

For each $u$ we thus get a collection of co-centric polygons. The integer lattice points lying on these correspond to holomorphic functions, see figure \ref{fig:PolygonPointsEx}. For example,  there are three holomorphic functions with $e_1$-charge $0$, $e_2$-charge $1$ and $R$-charge $7$ (explicitly, these functions are 
$z_1^2w_1^2z_2 z_3^2$, $z_2^3 w_2^2 z_3^2$, and $z_1w_1 z_2^2 w_2  z_3^2$, c.f. \eqref{eq:holofuns}). 

One way to write the contribution from $H^{(0,0)}_{\bar\partial_H}$ to the superdeterminant \eqref{eq:sdet} would be
\begin{align}
\prod\limits_{u=0}^\infty \prod\limits_{\substack{s,t,m,n\in \Z \\ m,n \geq 0 \\ |s|+|t|+|3s+2t|+2(m+n) = u}} \left( x + s\w_1 + t\w_2 + u \mu \right) \, ,
\end{align}
or for the general case:
\begin{equation} \label{eq:answerQt}
\prod\limits_{u=0}^\infty \prod\limits_{\substack{s,t,m,n\in \Z \\ m,n \geq 0 \\ \sum_i |\Qt_{i1}s+ \Qt_{i2} t|+2(m+n) = u}} \left(x + s\w_1 + t\w_2 + u \mu \right) \,.
\end{equation}
Here $\w_1, \w_2, \mu$ are equivariant parameters for $e_1, e_2$ and $R$ respectively.

\subsection{Holomorphic functions in terms of lattice points in a cone}\label{sec:Hfitolpiac}
We now turn to a geometric interpretation of previous enumeration result. This gives a systematic way of resolving the absolute values in \eqref{eq:ringrelations} and allows us to write the products above in terms of integer lattice points in a cone. 

Continuing from sec.\ref{sec:Hpa}, let $\{v_i\}$ be the normal vectors of the hyperplanes. As discussed in sec.\ref{sec:Hpa}, we have the freedom to pick $v_1=(1,0)$, $v_2=(0,1)$, and by re-ordering and flipping signs we may assume that they are ordered counter-clockwise.

For each $v_i$ we have constructed a vector $P_i$ as in sec.\ref{sec_Gctatfp}
\begin{equation}
P_i = v_i + \sum_{j \neq i} \sgn (\det [v_i,v_j]) v_j  \,. \nn
\end{equation}
See figure \ref{fig:normalvectorsR}.
The concentric polygons in the $s$-$t$ plane of the last section e.g. fig.\ref{fig:PolygonPointsEx} will have inward pointing normals
$\pm P_i$ (these polygons are all symmetric under inversion of $\BB{R}^2$).
If we think of those 2D polygons as viewing from the top a 3D cone $C$
with inward pointing normals
\begin{equation}
\vec{N}_i^\pm = (\pm P_i,1) \label{eq:Nidef}
\end{equation} 
The cone $C$ has its apex at the origin, see figure \ref{fig:cone1} for a sketch.

\begin{figure}[tbp]
    \centering
    \begin{minipage}[t]{0.45\textwidth}
        \centering
\begin{tikzpicture}[scale=0.75]
  \draw[thin,gray!40] (-4,-4) grid (4,4);
  \draw[->] (-4,0)--(4,0) node[right]{$s$};
  \draw[->] (0,-4)--(0,4) node[above]{$t$};
  \draw[line width=2pt,black,-stealth] (0,0)--(1,0) node[anchor=south west]{$\boldsymbol{v_1}$};
  \draw[line width=2pt,black,-stealth](0,0)--(0,1) node[anchor=north west]{$\boldsymbol{v_2}$};
    \draw[dashed,line width=2pt,gray,-stealth](4,-4)--(-4,4);
    \draw[line width=2pt,black,-stealth] (0,0)--(-1,1) node[] at (-0.75,1.25) (1) {$\boldsymbol{v_3}$};
    \draw[line width=2pt,black,-stealth] (0,0)--(-2,1) node[] at (-2.5,0.75) (1) {$\boldsymbol{v_4}$};
    \draw[line width=2pt,black,-stealth](0,0)--(-2,-3) node[anchor=north west] (2) {$\boldsymbol{v_{n-1}}$};
    \draw[line width=2pt,black,-stealth](0,0)--(3,-1) node[anchor=north west]{$\boldsymbol{v_{n}}$};
    \node[draw=none, black] at (-1.75,-0.25) {$\boldsymbol{\vdots}$};
    \node[draw=none, black] at (-1.5,-0.75) {$\ddots$};        		  \node[draw=none, black] at (-1.5,0.35) {\reflectbox{$\ddots$}};
    \node[draw=none, black] at (-1.5,2.5) {$\boldsymbol{-}$};
    \node[draw=none, black] at (-2.5,1.5) {$\boldsymbol{+}$};
\end{tikzpicture}
\caption{Construction of the $P_i$. In the example illustrated above we would get that $P_3 = v_3 -v_1 - v_2 + v_4 + \cdots +v_{n-1} - v_n$.}
\label{fig:normalvectorsR}
\end{minipage}\hfill
\begin{minipage}[t]{0.45\textwidth}
        \centering
\tdplotsetmaincoords{70}{0}
\begin{tikzpicture}[scale=0.75, 
 tdplot_main_coords,
 one end extended/.style={shorten >=-#1},
 one end extended/.default=1cm,
 ]
\def\RI{3}
\def\RII{1.25}
\def\kI{7}
\def\kII{3}
\draw [line width=0.25ex, color=gray] (0,\kI/3) node at (0,\kI/3) (labe1kI1) {} -- (\kI/4,0) node at (\kI/4,0) (labe1kI2) {}  -- (2/5*\kI,-3/5*\kI) node at (2/5*\kI,-3/5*\kI) (labe1kI3) {}  --(0,-\kI/3) node at (0,-\kI/3) (labe1kI4) {}  -- (-\kI/4,0)  node at (-\kI/4,0) (labe1kI5) {} -- (-2/5*\kI,3/5*\kI)  node at (-2/5*\kI,3/5*\kI) (labe1kI6) {} -- cycle;
\begin{scope}[yshift=-4cm]
\node[draw=none, blue] at (0,0) (origin) {};
\end{scope}
\foreach \i in {1,2,3,4,5,6}
{
\draw [one end extended=2cm] (origin.center) -- (labe1kI\i.center);
}
\end{tikzpicture}
\caption{Sketch of the 3D cone $C$.}
\label{fig:cone1}
    \end{minipage}
\end{figure}

We then consider integer lattice points inside this cone but with some caveats:
Firstly, only `half' of the points will be included. We only include the points $\vec{p}=(s,t,u)$ that satisfy (c.f. \eqref{eq:muweight})
\begin{equation}
\vec{N}_i^\pm \cdot \vec{p} \equiv 0 \pmod{2} \label{eq:mod2constraint} \, .
\end{equation}
Noting that all $N^\pm_i$ are congruent to each other modulo 2, we 
define the 3-vector
\begin{equation}
\vec{N} = \left( \sum_i v_i, 1 \right ) \pmod 2 \,, \label{eq:Nvecdef}
\end{equation}
where we take the modulo 2 reduction of each component, and instead state condition \eqref{eq:mod2constraint} as
\begin{equation}
\vec{N} \cdot \vec{p} \equiv 0 \pmod{2} \label{eq:mod2constraintN} \, .
\end{equation}
Secondly, the points in the cone come with multiplicities. The points on the face of the cone have multiplicity one, and then the multiplicities increase as one goes further inside the cone. Concretely the multiplicity determined by the distance to the face of the cone, appropriately scaled, and is given by 
\begin{equation}
d(\vec{p},\partial C) = \frac{1}{2}\min_i \vec{N}_i^\pm \cdot \vec{p}+1 \,. \label{eq:multiplicity}
\end{equation}
Due to \eqref{eq:mod2constraint} this is always an integer. The minimal $\min_i \vec{N}_i^\pm \cdot \vec{p}$ is realised by a different face depending on the region one is in. This corresponds to the region by region resolution of the absolute value in the last section.

Putting this together, we enumerate the holomorphic functions and find the $H^{(0,0)}_{\bar\partial_H}$-contribution to the superdeterminant \eqref{eq:sdet} to be (where $\vec p=(s,t,u)$):
\begin{equation}
\prod \limits_{\substack{\vec{p} \in C \cap \Z^3 \\ \vec{p}\cdot \vec{N} \equiv 0 \pmod{2}}}
(x + s \w_1 + t \w_2 + u \mu )^{d(\vec{p},\partial C)} \label{eq:sdetH0contrib} \, .
\end{equation}
This expression is equivalent to \eqref{eq:answerQt} of the previous subsection.

We have already explained in sec.\ref{sec:one-loop} how to deal with the $H^{(0,3)}_{\bar\partial_H}$-contribution to the super-determinant. It is obtained by changing the sign of the weights and shifting the $R$-weight:
\begin{equation}
\prod \limits_{\substack{\vec{p} \in C \cap \Z^3 \\ \vec{p}\cdot \vec{N} \equiv 0 \pmod{2}}}
(x - s \w_1 - t \w_2 - u \mu -4\mu)^{d(\vec{p},\partial C)} \label{eq:sdetH3contrib} \,.
\end{equation}
We combine these two contributions together and define a special function
\begin{equation}
\boxed{S^C(x|\w_1,\w_2,\mu) =  \frac{\prod \limits_{\substack{\vec{p} \in C \cap \Z^3 \\ \vec{p}\cdot \vec{N} \equiv 0 \pmod{2}}}
(x + s \w_1 + t \w_2 + u \mu )^{d(\vec{p},\partial C)} }{\prod \limits_{\substack{\vec{p} \in C^{\circ} \cap \Z^3 \\ \vec{p}\cdot \vec{N} \equiv 0 \pmod{2}}}
(-x + s \w_1 + t \w_2 + u \mu )^{d(\vec{p},\partial {C})-2}}}  \label{eq:SQdef} \, ,
\end{equation}
where $C$ (resp. $C^{\circ}$) is the (resp. interior of the) cone with apex at the origin and inward pointing normals given by \eqref{eq:Nidef}. The multiplicities $d(\vec{p},\partial C)$ are defined in \eqref{eq:multiplicity}. 
Note, in comparison with \eqref{eq:sdetH3contrib}, we flipped the sign of the factors in the denominator. This is for the sake of using zeta function regulation to regulate the infinite product. 

In summary the super-determinant is concisely
\begin{equation}
\sdet_{\Omega_{KR}^{(0,\bullet)}} (-\L_R + x)=S^C(x|\go_1,\go_2,\mu) \,.
\end{equation}
The perturbative partition function for maximally supersymmetric Yang-Mills theory on this manifold is
\begin{equation}
Z^{pert} = \int\limits_t d \sigma \,  e^{-\frac{24}{g^2_7} V_7 \Tr(\sigma^2)} \prod_{\beta \neq 0} \big|S^C( i \bra \sigma, \beta \ket |\w_1,\w_2,\mu)\big|\, , \label{pertparfunQ}
\end{equation}
very much analogous to the toric Sasaki case reviewed in sec.\ref{sec:one-loop}.  Note even though the 3-Sasaki geometry corresponds to $\go_1=\go_2=0$ (unrefined), we have kept the $\go_{1,2}$ in the above result as a refined partition function. It is not clear to us what is the deformation of 3-Sasaki geometry that corresponds to setting $\go_{1,2}\neq 0$ in the Reeb. So our refinement is done formally, as opposed to the 5D toric Sasaki case where the deformation of the Reeb is well-understood. This also affects our computation of the refined volume undertaken in sec.\ref{sec.Tft}.

\subsection{An example in \cite{Herbig2014}}
Let us now turn to the previously discussed example whose $\Qt$-matrix is given by \eqref{eq:Qtmatrix321} and whose hyperplanes are illustrated in figure \ref{fig:hyperplaneregions}. We have the hyperplane normals
\begin{align}
v_1 &= (1,0) \, , \nn \\
v_2 &= (0,1) \,, \label{eq:321hyperplanenormals} \\
v_3 &= (3,2) \, , \nn 
\end{align}
and from \eqref{eq:Rdef} we find
\begin{align}
P_1 &= v_1 + v_2 + v_3 = (4,3) \, , \nn \\
P_2 &= v_2 - v_1 - v_3 = (-4,-1) \,, \label{eq:321Rvectors} \\
P_3 &= v_3 - v_1 + v_2 = (2,3) \,. \nn 
\end{align}                                                                                                                                                                                                                                                                                                                                                                                                                                                                                                                                                                                                                                                                                                                                                                                                                                                                                                                                                                                                                                                                                                                                                                                                                                                                                                                                                                                                                                                                                                                                                                                                                                                                                                                                                                                                                                                                                                                                                                                                                                                                                                                                                                                                                                                                                                                                                                                                                                                                                                                                                                                                                                                                                                                              
                                                                                                                                                                                                                                                                                                                                                                                                                                                                                                                                                                                                                                                                                                                                                                                                                                                                                                                                                                                                                                                                                                                                                                                                                                                                                                                                                                                                                                                                                                                                                                                                                                                                                                                                                                                                                                                                                                                                                                                                                                                                                                                                                                                                                                                                                                                                                                                                                                                                                                                                                                                                                                                                                                                              We thus consider the 3D cone $C$ with apex at the origin and inward-pointing normals
                                                                                                                                                                                                                                                                                                                                                                                                                                                                                                                                                                                                                                                                                                                                                                                                                                                                                                                                                                                                                                                                                                                                                                                                                                                                                                                                                                                                                                                                                                                                                                                                                                                                                                                                                                                                                                                                                                                                                                                                                                                                                                                                                                                                                                                                                                                                                                                                                                                                                                                                                                                                                                                                                                                             \begin{align}                                                                                                                                                                                                                                                                                                                                                                                                                                                                                                                                                                                                                                                                                                                                                                                                                                                                                                                                                                                                                                                                                                                                                                                                                                                                                                                                                                                                                                                                                                                     \vec{N}_1^+ &= (P_1,1) = (4,3,1) \,, & \vec{N}_1^- &= (-P_1,1) = (-4,-3,1)  \,,\nn \\
\vec{N}_2^+ &= (P_2,1) = (-4,-1,1) \,, & \vec{N}_2^- &= (-P_2,1) = (4,1,1)  \,,  \\
\vec{N}_3^+ &= (P_3,1) = (2,3,1) \,, & \vec{N}_3^- &= (-P_3,1) = (-2,-3,1)  \,. \label{N_example}
\end{align}
%and the cone $\hat{C}$ with the same normals but whose apex is at $(0,0,4)$.
The vector $\vec{N}$ in \eqref{eq:Nvecdef} is given by $\vec{N} = (0,1,1) $ and the condition \eqref{eq:mod2constraintN} is thus equivalent to $p_2 \equiv p_3 \pmod{2}$.
We can thus write the function \eqref{eq:SQdef} easily with the data from \eqref{N_example}.

To explicitly enumerate the integer lattice points in such a cone $C$ requires some work. For some simple examples one can do this by hand. But for more general examples resort to computer help to find the lattice points.
%\rbox{A Matlab/Mathematica script to do so can be obtained from the authors upon request.}

Above we have written expressions involving the equivariant parameters $\w_1$ and $\w_2$, but we can also ignore these to get an `unrefined' answer, which was done in \cite{Herbig2014}. If we set $\go_i=0$ the numerator in \eqref{eq:SQdef} for our current example and just count the total multiplicity for each $\mu$-coefficient we obtain the following multiplicities:
\begin{center}
\begin{tabular}{|c|cccccccccccc|}
\hline
$\mu$-coefficient :   & 0 & 1 & 2 & 3 & 4 & 5 & 6 & 7 & 8 & 9 & 10 & $\cdots$\\
\hline
multiplicity: & 1 & 0 & 2 & 4 & 7 & 10 & 16 & 22 & 31 & 40 & 54 & $\cdots$\\
\hline
\end{tabular}
\end{center}
The sequence of multiplicities is A244488 in OEIS \cite{OEIS} and corresponds to the `Hilbert series' of the HK cone as computed in \cite{Herbig2014}, The generating function is given by:
\begin{equation}
\frac{1+x^2+3x^3+4x^4+4x^5+4x^6+3x^7+x^8+x^{10}}{(1-x^2)(1-x^3)(1-x^4)(1-x^5)} = 1 + 2x^2 + 4 x^3 + 7 x^4 + 10x^5 + 16x^6 + \cdots
\end{equation}
We see here the explicit match with our result. Our calculations thus provide a combinatorial interpretation of these coefficients in terms of integer lattice points inside cones.

\section{Factorisation: examples} \label{sec:factorisation}
We now turn to studying factorisation properties of the function $S^C$ defined in \eqref{eq:SQdef}. The factorisation is motivated by physics consideration. Indeed, the 7D 3-Sasakian manifolds are a fibration over some 4D orbifold $B$ with \emph{generic} fibre $S^3$ or $SO(3)$
 \bea
  \begin{tikzpicture}
  \matrix (m) [matrix of math nodes, row sep=1.4em, column sep=1.6em]
    { X^7 & S^3{~\rm or}~SO(3)  \\
    B &   \\ };
  \path[->]
  (m-1-2) edge (m-1-1)
  (m-1-1) edge (m-2-1);
 \end{tikzpicture}.\nn\eea
The fibre can degenerate at certain orbifold points of $B$:  
The ${\mathbb T}^2$ acting on $X^7$ descends to an action on $B$. Over these fixed points the fibre degenerates. For the hyper-toric case we study, the degeneration will be  of type $\Gc\backslash S^3$ for some cyclic group $\Gc\subset SU(2)$, see sec.13.3.5 in \cite{BoyerGalicki}. We have shown in sec.\ref{sec_Gctatfp} that the neighbourhood of such a fibre has geometry ${\cal S}(r_i)$ i.e. $\Gc=\BB{Z}_{r_i}$. The normal bundle of the fibre splits as ${\cal O}(-q_i)\oplus{\cal O}(q_i-2)$, where $r_i$ and $q_i$ are defined below \eqref{eq:Rdef}.

For us it is important that the ${\mathbb T}^2$ would act on the instanton moduli space ${\cal M}$ (which we do not yet have). The fixed points of the ${\mathbb T}^2$ action on ${\cal M}$ should correspond to instantons supported at the fibres sitting above the ${\mathbb T}^2$-fixed points in $B$. This is in analogy with the 5D situation where the 
torus invariant instantons are point like and propagates along the isolated closed Reeb orbits. Here the closed Reeb orbits are replaced with $\Gc\backslash S^3$.

From the experience of 5D Nekrasov instanton partition functions, the 7D partition function should also be obtained by stitching together partition functions on $\Gc\backslash S^3\times_{\rm tw}\BB{C}^2$, where $\times_{\rm tw}$ means twisted product, since, as we have seen, the normal bundle (the $\BB{C}^2$) is fibred over $\Gc\backslash S^3$, Also one needs to take into account possible fluxes, i.e. non-zero $c_1$ of the gauge bundle supported on torus invariant 4-cycles.

This is a generalisation of the similar results for 5D SYM on toric SE manifolds  \cite{Qiu:2014oqa}, where it was shown that the perturbative partition function are factorised into perturbative Nekrasov partition functions \cite{Nekrasov:2008kza} on $S^1 \times \C^2$. 
In coming sections we show that the 7D perturbative part indeed has this pattern, and the constituent factor corresponding to $\Gc\backslash S^3\times_{\rm tw} \BB{C}^2$ does exhibit features of the representation ring of $SU(2)$ acting on $\Gc\backslash S^3$.

%For SYM on 7D toric Sasaki-Einstein manifolds one obtains similar factorisations in terms of Nekrasov partition functions on $S^1 \times \C^3$, however the grounds for conjecturing the full answer from this factorisation are weaker, see \cite{Polydorou2017} for a discussion.

\subsection{A second look at $S^7$}
It is instructive to take a second look at $S^7$ which is toric Sasaki-Einstein as well as hyper-toric 3-Sasakian. It as two fibration pictures fig.\ref{fig:Hopffibrations}.
\begin{figure}[h]
\centering
\begin{tikzcd}
S^7  \arrow[d] & S^1 \arrow[l] & & & S^7  \arrow[d] & S^3 \arrow[l]\\
\C P ^3  & & & & S^4 \iso \mathbb{H}P^1  & 
\end{tikzcd}
\caption{Complex and quaternionic Hopf fibrations of $S^7$, the $S^1$ fibre on the left corresponds to the Hopf of the $S^3$ on the right.}\label{fig:Hopffibrations}
\end{figure}

If one computes the perturbative partition function by treating $S^7$ as a toric SE manifold, one obtains the quadruple sine function $S_4$ \cite{Polydorou2017}. The function $S_4$ is well-known to be factorisable into four copies of q-factorials, which in turn corresponds to the local geometry $S^1\times\BB{C}^3$. Here $\BB{C}^3$ is the neighbourhood of one of the four torus fixed points on $\BB{C}P^3$ and $S^1$ is the Hopf fibre sitting on the fixed point.
All of this corresponds to the complex Hopf fibration in the figure. 

We can also view $S^7$ as an $S^3$-bundle over $S^4$, corresponding to the quaternionic Hopf fibration. It is natural to ask if this fibration gives rise to factorisations similar to the Sasaki-Einstein case, but exhibiting a different facet of the geometry.

We start with viewing $S^7$ using the left picture of fig.\ref{fig:Hopffibrations}.
One obtains a quadruple sine function in the perturbative partition function \cite{Minahan:2015jta, Polydorou2017}. Recall that the infinite product expression for the quadruple sine is given by
\begin{equation}
S_4 (x|\wt_1,\wt_2,\wt_3,\wt_4) = \frac{\prod\limits_{i,j,k,l = 0}^\infty \left(x+ i\wt_1+j\wt_2+k\wt_3+l\wt_4 \right) } {\prod\limits_{i,j,k,l = 1}^\infty \left(-x +i\wt_1+j\wt_2+k\wt_3+l\wt_4  \right)} \, ,
\label{eq:S4Defn}
\end{equation}
where the numerator is easily seen as the equivariant enumeration of holomorphic functions of $\BB{C}^4$, with $\wt_{1,2,3,4}$ being the equivariant parameters for the four $U(1)$'s acting on $\BB{C}^4$ in the standard manner.

If we instead view $S^7$ as hypertoric 3-Sasakian, the two U(1)'s denoted as $e_1,e_2$ in sec.\ref{sec:counting} act with weights $(1,0,-1,0)$ and $(0,1,0,-1)$. %This corresponds to fixing $\wt_3=-\wt_1$ and $\wt_4=-\wt_2$. The equivariant parameter for the Reeb ($\wt_1+\wt_2+\wt_3+\wt_4$) would then be zero and we see that \eqref{eq:S7SQfun} and \eqref{eq:S4Defn} match.
%the product in the numerator of \eqref{eq:S4Defn} reads
%\bea \prod\limits_{i,j,k,l = 0}^\infty \left(x+ i\wt_1+j\wt_2+k\wt_3+l\wt_4 \right)=\prod\limits_{i,j,k,l = 0}^\infty \left(x+ (i-k)\w_1+(j-l)\w_2+(i+j+k+l)\mu \right).\nn\eea
From the general treatment described in sec.\ref{sec:Hfitolpiac}
the hyperplane arrangement for $C(S^7) = \C^4$ is shown in figure \ref{fig:hyperplanesS7} and the two normals are
\begin{align}
v_1 &= (1,0) \,, \\
v_2 &= (0,1) \,.
\end{align}
According to the prescription
\begin{align}
P_1 &=  v_1 + v_2 = (1,1) \, , \\
P_2 &=  v_2 - v_1 = (-1,1) \, , 
\end{align}
and the four normals of our polygons are thus 
\begin{align}
\vec{N}_1^+ &= (P_1,1) = (1,1,1) \, , \\
\vec{N}_1^- &= (-P_1,1) = (-1,-1,1) \, , \\
\vec{N}_2^+ &= (P_2,1) = (-1,1,1) \, , \\
\vec{N}_2^- &= (-P_2,1) = (1,-1,1) \, .
\end{align}
For a fixed height in the $(0,0,1)$-direction we get co-centric squares as in figure \ref{fig:PolygonPointsS7}.
\noindent
\begin{figure}[tbp]
\centering
\noindent
\begin{minipage}[t]{0.45\textwidth}
\noindent
\centering
\noindent
\begin{tikzpicture} [scale=0.7]
\draw [->] (-5,0) -- (5,0) node [below] {\small $s$};
\draw [->] (0,-5) -- (0,5) node [left] {\small $t$};
\draw [line width=0.5ex] (-4,0) -- (4,0);
\draw [line width=0.5ex] (0,-4) -- (0,4);
\foreach \x in {-4,...,4}
\foreach \y in {-4,...,4}
	{
	\draw [black,fill=black] (\x,\y) circle (0.5ex);
	 }
\node at (3.5,3.5) {\large I};
\node at (3.5,-3.5) {\large II};
\node at (-3.5,-3.5) {\large III};
\node at (-3.5,3.5) {\large IV};
\end{tikzpicture}
\caption{Hyperplanes for $C(S^7)=\C^4$.}
\label{fig:hyperplanesS7}
\end{minipage}
\hfill
\noindent
\begin{minipage}[t]{0.45\textwidth}
\noindent
\centering
\noindent
\begin{tikzpicture}[scale=.7]
\draw [->] (-5,0) -- (5,0) node [below] {\small $s$};
\draw [->] (0,-5) -- (0,5) node [left] {\small $t$};
\draw [line width=0.5ex] (-4,0) -- (4,0);
\draw [line width=0.5ex] (0,-4) -- (0,4);
\foreach \x in {-4,...,4}
\foreach \y in {-4,...,4}
	{
	\draw [black,fill=black] (\x,\y) circle (0.5ex);
	 }
\foreach \k in {0,2,4}
	{\draw [line width=0.25ex, color=black] (0,\k) -- (\k,0) -- (0,-\k) --(-\k,0)-- cycle;
	}
\node at (3.5,3.5) {\large I};
\node at (3.5,-3.5) {\large II};
\node at (-3.5,-3.5) {\large III};
\node at (-3.5,3.5) {\large IV};
\node[circle,scale=0.65,color=white, fill=black] at (0,0) {$\boldsymbol{3}$};
\foreach \k in {0,...,4}
	{
	\node[circle,scale=0.65,color=white, fill=black] at (\k,4-\k){$\boldsymbol{1}$};
	\node[circle,scale=0.65,color=white, fill=black] at (-\k,-4+\k){$\boldsymbol{1}$};
	}
\foreach \k in {1,2,3}
	{
	\node[circle,scale=0.65,color=white, fill=black] at (-4+\k,\k){$\boldsymbol{1}$};
	\node[circle,scale=0.65,color=white, fill=black] at (-\k+4,-\k){$\boldsymbol{1}$};
	\node[circle,scale=0.65,color=white, fill=black] at (3-\k,\k-1) {$\boldsymbol{2}$};
	\node[circle,scale=0.65,color=white, fill=black] at (-3+\k,-\k+1) {$\boldsymbol{2}$};
	}
\node[circle,scale=0.65,color=white, fill=black] at (-1,1) {$\boldsymbol{2}$};
\node[circle,scale=0.65,color=white, fill=black] at (1,-1) {$\boldsymbol{2}$};
\end{tikzpicture}
\caption{Slice of cone for $S^7$ where $u=4$.}
\label{fig:PolygonPointsS7}
\end{minipage}
\end{figure}

We count integer lattice points inside the cone with apex at the origin and with inward-pointing normals given above. Only the points satisfying \eqref{eq:mod2constraintN} are included, i.e. the points $\vec{p} = (s,t,u)$ satisfying 
\begin{equation}
s + t + u \equiv 0 \pmod{2} \,,
\end{equation}
and their multiplicities are given by \eqref{eq:multiplicity}
\begin{equation}
d(\vec{p},\partial C) = \frac{\min_i \vec{N}_i^\pm \cdot \vec{p}}{2}+1  = \min  \frac{\pm s \pm t + u}{2}+1  \,. \label{eq:multiplicityS7}
\end{equation}

The cone in this example is easy to  describe and we can enumerate the integer lattice points  satisfying $s + t + u \equiv 0 \pmod{2}$, with the correct multiplicities, via
\begin{align}
i(1, 0, 1)+j(0,1,1)+k(-1,0,1)+l(0,-1,1),~~i,j,k,l=0,1, \dots \, .
\end{align}
The function $S^C$ in \eqref{eq:SQdef} for $S^7$ can then be written as
\begin{equation}
S^{C(S^7)}(x|\w_1,\w_2,\mu) =  \frac{\prod \limits_{i,j,k,l=0}^\infty (x + (i-k)\w_1 + (j-l)\w_2 + (i+j+k+l) \mu )}{\prod \limits_{i,j,k,l=1}^\infty (-x + (i-k)\w_1 + (j-l)\w_2 + (i+j+k+l) \mu )} \,. \label{eq:S7SQfun}
\end{equation}
One sees that this coincides with \eqref{eq:S4Defn} if one makes the identification $\tilde\go_1=\go_1+\mu$, $\tilde\go_3=-\go_1+\mu$, $\tilde\go_2=\go_2+\mu$ and $\tilde\go_4=-\go_2+\mu$.

\smallskip
We turn next to factorisation that reflects the $S^3$ fibration. 
It is by now a standard result from localisation calculations (amongst which \cite{KaWiYa} is the earliest) of Chern-Simons theory that a 3-sphere should contribute a double sine function $S_2$. Our
alternative factorisation of $S_4$ into $S_2$ corresponds simply to reorganising the product. But we do point out that the infinite products are all defined under zeta function regularisation, therefore any such reorganisation would produce certain Bernoulli polynomials. But for this section, we put these aside and proceed most naively. 

For each hyperplane we will get a factor in the form of an infinite product of double sine functions. Recall the double sine function is given by the infinite product\footnote{where zeta function regularisation is implicitly understood.}
\begin{equation}
S_2(x|\w_1,\w_2) = \frac{\prod \limits_{k,l=0}^{\infty} (x+k\w_1+l w_2)}{\prod \limits_{k,l=1}^{\infty}(-x+k\w_1+l w_2)} \,. \label{eq:S2ordinarydef}
\end{equation}
More generally for cone of dimension 2 the generalised double sine is defined by \cite{Winding:2016wpw}
\begin{equation}
S_2^C(x|\vec{\w}=(\w_1,\w_2)) = \frac{\prod \limits_{\vec{n} \in C \cap \Z^2} (x+\vec{n} \cdot \vec{\w})}{\prod \limits_{\vec{n} \in C^{\circ} \cap \Z^2} (-x+\vec{n} \cdot \vec{\w})} \,.
\end{equation}
Here $C^\circ$ denotes the interior of the cone. Taking $C=\R^2_{\geq 0}$ revert to the previous double sine.

For the hyperplane with $v_1$, we construct the double sine
\bea S_2(x+(a-b)\go_1-(a+b+1)\go_2-\mu|\mu+\go_2,\mu-\go_2),~~a,b\in\BB{Z}_{\geq0}\nn\eea
while for the hyperplane with $v_2$, we construct the double sine
\bea S_2(x-(a+b+1)\go_2+(b-a-1)\mu|\mu-\go_1,\mu+\go_1)^{-1},~~a,b\in\BB{Z}_{\geq0}.\nn\eea
We claim that 
\bea S^{C(S^7)}\sim \frac{\prod_{a,b\geq0}S_2(x+(a-b)\go_1-(a+b+1)\go_2-\mu|\mu+\go_2,\mu-\go_2)}{\prod_{a,b\geq0}S_2(x-(a+b+1)\go_2+(b-a-1)\mu|\mu-\go_1,\mu+\go_1)}.\label{eq:fact_S7_simple}\eea
The sign $\sim$ means up to Bernoulli factors. Practically, this means that the two sides will have the same pole and zeros, but their asymptotic behaviour can disagree (the missing Bernoulli polynomial will eventually correct this).

Let us see a few checks. From fig.\ref{fig:PolygonPointsS7} the $S^{C(S^7)}$ should have a triple zero at $x+4\mu=0$ and simple pole at $x-4\mu=0$. As for $S_2(x|u,v)$, it has a zero at $x+pu+qv=0$ and a pole at $x-(p+1)u-(q+1)v=0$ for $p,q\geq0$. In \eqref{eq:fact_S7_simple}, the numerator has a zero at
\bea x+(a-b)\go_1-(a+b+1)\go_2-\mu+p(\mu+\go_2)+q(\mu-\go_2)=0,~a,b,p,q\geq0.\nn\eea
We can set $(a,b,p,q)=(0,0,3,2),\,(1,1,4,1),\,(2,2,5,0)$ to get a zero at $x+4\mu=0$. Similarly only at $(a,b,p,q)=(0,0,0,1)$ does the numerator give a pole $x-4\mu=0$.
The denominator contains $-(a+b+1)\go_2$ and will never contribute pole or zero at $x\pm4\mu=0$. Thus the two sides match.
Similarly $S^{C(S^7)}$ has a zero at $x-3\go_1+\go_2+4\mu=0$, matched the same zero from the $S_2$ in the numerator for $(a,b,p,q)=(0,3,5,0)$.

A more interesting check is that $S^{C(S^7)}$ has no zero at, say, $x+2\go_1-4\go_2+4\mu=0$, or pole at $x-2\go_1-4\go_2-4\mu=0$, as seen from fig.\ref{fig:PolygonPointsS7}. But 
the numerator of \eqref{eq:fact_S7_simple} has a quadruple zero at $x+2\go_1-\go_2+4\mu=0$. But this is entirely artificial, because the denominator has the same quadruple zero. In the same way, the numerator has a spurious double pole $x-2\go_1-4\go_2-4\mu=0$, which gets cancelled by the same double pole from the denominator. 

It is natural to wander what is the systematics in these fortuitous cancellations. In fact the rewriting \eqref{eq:fact_S7_simple} comes from computing the equivariant index of $\bar\partial_H$ from localisation, where the two $S_2$ factors come from the 3-sphere sitting at the north and south pole of the base 4-sphere. But the index calculation will be presented elsewhere.
 
\subsection{The Swann bundle}
The next example is the 3-Sasakian manifold $X$ associated to the Swann bundle. We first use the machinery of sec.\ref{sec:Hfitolpiac} to redo the computation done in \cite{Rocen2018}, but in much more streamlined fashion.

\begin{figure}[tbp]
\centering
\begin{tikzpicture} [scale=0.7]
\draw [->] (-5,0) -- (5,0) node [below] {\small $s$};
\draw [->] (0,-5) -- (0,5) node [left] {\small $t$};
\draw [line width=0.5ex] (-4,0) -- (4,0);
\draw [line width=0.5ex] (0,-4) -- (0,4);
\draw [line width=0.5ex] (4,-4) -- (-4,4);
\foreach \x in {-4,...,4}
\foreach \y in {-4,...,4}
	{
	\draw [black,fill=black] (\x,\y) circle (0.5ex);
	 }
%\node at (3.5,3.5) {\large I};
%\node at (3.5,-3.5) {\large II};
%\node at (-3.5,-3.5) {\large III};
%\node at (-3.5,3.5) {\large IV};
\end{tikzpicture}
\caption{Hyperplanes for $C(\S)$.}
\label{fig:hyperplanesSwann}
\end{figure}
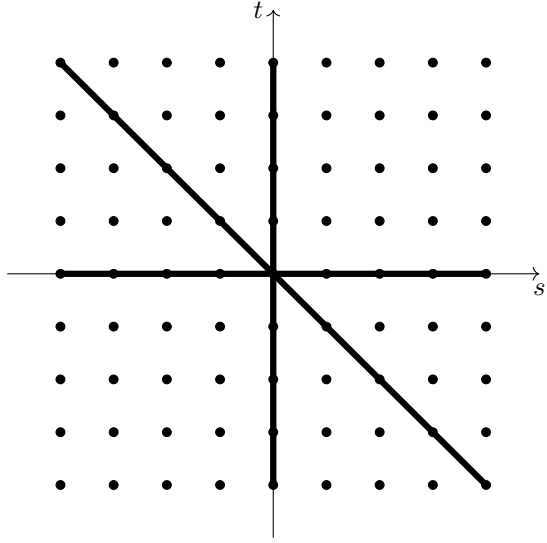

The hyperplanes for the cone $C(X)$ are illustrated in figure \ref{fig:hyperplanesSwann} and their normals are given by
\begin{align}
v_1 &= (1,0) \,, \\
v_2 &= (0,1) \,, \\
v_3 &= (1,1) \,.
\end{align}
From these normals we calculate
\begin{align}
P_1 &=  (2,2) \, , \\
P_2 &=  (-2,0) \, ,  \\
P_3 &=  (0,2) \, , 
\end{align}
and
\begin{align}                                                                                                                                                                                                                                                                                                                                                                                                                                                                                                                                                                                                                                                                                                                                                                                                                                                                                                                                                                                                                                                                                                                                                                                                                                                                                                                                                                                                                                                                                                                     \vec{N}_1^+ &= (2,2,1) \,, & \vec{N}_1^- &= (-2,-2,1)  \,,\nn \\
\vec{N}_2^+ &= (-2,0,1) \,, & \vec{N}_2^- &= (2,0,1)  \,,  \\
\vec{N}_3^+ &= (0,2,1) \,, & \vec{N}_3^- &= (0,-2,1)  \,.\nn
\end{align}
The mod 2 vector $\vec{N}$ is
\begin{equation}
\vec{N} = (0,0,1) \,,
\end{equation}
which tells us that the fibre is SO(3). We will only get even $\mu$-weights $u$ and there are no further restrictions on the $(s,t)$-coordinates. The polygons for a fixed $u$ are illustrated in figure \ref{fig:PolygonPointsSwann}.

\begin{figure}[tbp]
\centering
\begin{minipage}[t]{0.45\textwidth}
\centering
\begin{tikzpicture}[scale=.6]
\draw [->] (-5,0) -- (5,0) node [below] {\small $s$};
\draw [->] (0,-5) -- (0,5) node [left] {\small $t$};
\draw [line width=0.25ex] (-4,0) -- (4,0);
\draw [line width=0.25ex] (0,-4) -- (0,4);
\draw [line width=0.25ex] (4,-4) -- (-4,4);
\foreach \x in {-4,...,4}
\foreach \y in {-4,...,4}
	{
	\draw [black,fill=black] (\x,\y) circle (0.5ex);
	 }
\foreach \k in {0,1,2}
	{\draw [line width=0.25ex, color=black] (-\k,\k) -- (0,\k) -- (\k,0) --(\k,-\k) -- (0,-\k) -- (-\k,0) -- cycle;
	}
\node[circle,scale=0.65,color=white, fill=black] at (0,0) {$\boldsymbol{3}$};
\foreach \k in {-1,1}
{
	\node[circle,scale=0.65,color=white, fill=black] at (0,\k) {$\boldsymbol{2}$};
	\node[circle,scale=0.65,color=white, fill=black] at (\k,0) {$\boldsymbol{2}$};
	\node[circle,scale=0.65,color=white, fill=black] at (-\k,\k) {$\boldsymbol{2}$};
	\node[circle,scale=0.65,color=white, fill=black] at (\k,\k) {$\boldsymbol{1}$};
}
\foreach \k in {-2,2}
{
	\node[circle,scale=0.65,color=white, fill=black] at (0,\k) {$\boldsymbol{1}$};
	\node[circle,scale=0.65,color=white, fill=black] at (\k,0) {$\boldsymbol{1}$};
	\node[circle,scale=0.65,color=white, fill=black] at (-\k,\k) {$\boldsymbol{1}$};
}
\node[circle,scale=0.65,color=white, fill=black] at (-2,1) {$\boldsymbol{1}$};
\node[circle,scale=0.65,color=white, fill=black] at (-1,2) {$\boldsymbol{1}$};
\node[circle,scale=0.65,color=white, fill=black] at (2,-1) {$\boldsymbol{1}$};
\node[circle,scale=0.65,color=white, fill=black] at (-1,2) {$\boldsymbol{1}$};
\node[circle,scale=0.65,color=white, fill=black] at (1,-2) {$\boldsymbol{1}$};
\end{tikzpicture}
\caption{A slice of the cone at $u=4$ corresponding to the example $\S$ discussed in the text.}
\label{fig:PolygonPointsSwann}
\end{minipage}
\hfill
\begin{minipage}[t]{0.45\textwidth}
\centering
\begin{tikzpicture}[scale=.6]
\draw [->] (-5,0) -- (5,0) node [below] {\small $s$};
\draw [->] (0,-5) -- (0,5) node [left] {\small $t$};
\foreach \k in {0,1,2}
	{\draw[fill=gray!50,opacity=.6] (-\k,\k) -- (-\k+2,\k) -- (-\k+2,\k-2) -- (-\k,\k-2) -- cycle;}
\foreach \k in {0,1,2}
	{\draw[line width=1pt](-\k,\k) -- (-\k+2,\k) -- (-\k+2,\k-2) -- (-\k,\k-2) -- cycle;}
\draw [line width=0.25ex] (-4,0) -- (4,0);
\draw [line width=0.25ex] (0,-4) -- (0,4);
\draw [line width=0.25ex] (4,-4) -- (-4,4);
\foreach \x in {-4,...,4}
\foreach \y in {-4,...,4}
	{
	\draw [black,fill=black] (\x,\y) circle (0.5ex);
	 }
\end{tikzpicture}
\caption{The points in such a slice can be enumerated by considering overlapping squares.}
\label{fig:PolygonPointsSwannSquares}
\end{minipage}
\end{figure}

To make a connection with the result in \cite{Rocen2018} we can enumerate these lattice points as follows: For a fixed $u$, define $n=u/2$ (recall that $u$ is always even for this example). We can enumerate the lattice points with the correct multiplicities as follows:
\begin{align}
(k-i, j-k, 2n) \,, \quad 0\leq i,j,k \leq n \,.
\end{align}
This corresponds to covering the polygon by overlapping squares as illustrated in figure \ref{fig:PolygonPointsSwannSquares}.

We get
\begin{equation}
S^{C(X)}(x|\w_1,\w_2,\mu) =  \frac{\prod \limits_{n=0}^\infty \prod \limits_{i,j,k=0}^{n} (x + (k-i)\w_1 + (j-k)\w_2 + 2n \mu)}{\prod \limits_{n=0}^\infty \prod \limits_{i,j,k=0}^{n} (-x + (k-i)\w_1 + (j-k)\w_2 + 2(n+2) \mu )} \,, \label{eq:SwannSQfun}
\end{equation}
which is equivalent to the function obtained in \cite{Rocen2018}.

For the `unrefined' case, i.e. setting $\go_{1,2}=0$, this function can be written as 
\begin{equation}
S^{C(X)}(x) =  \frac{\prod \limits_{n=0}^\infty (2n + x)^{(n+1)^3}}{\prod \limits_{n=2}^\infty (2n - x)^{(n-1)^3}} \,.
\end{equation}
The numerator matches the Hilbert series given in \cite[section 5.3]{Herbig2014}:
\begin{equation}
\frac{1+4x^2+x^4}{(1-x^2)^4} = \sum_{n=0}^\infty (n+1)^3 x^{2n} \,.
\end{equation}

\smallskip

We turn to factorisation next. Since $X$ is a fibration over $\BB{C}P^2$
\begin{figure}[h]
\centering
\begin{tikzpicture}
  \matrix (m) [matrix of math nodes, row sep=1.4em, column sep=1.6em]
    { X & SO(3)  \\
    \BB{C}P^2 &   \\ };
  \path[->]
  (m-1-2) edge (m-1-1)
  (m-1-1) edge (m-2-1);
\end{tikzpicture}
\end{figure}
and there are three torus fixed loci with fibre $SO(3)$. That the fibre is $SO(3)$ also implies certain changes to their contributions: we define a 'half double sine' function
\begin{equation}
\hat S_2(x|\w_1,\w_2) = \frac{\prod \limits_{k,l\geq0,k+l={\rm odd}} (x+k\w_1+l w_2)}{\prod \limits_{k,l\geq1,k+l={\rm odd}}^{\infty}(-x+k\w_1+l w_2)} \,. \label{eq:S2halfdef}
\end{equation}
To $v_1,v_2,v_3$ associate 
\bea v_1:&& I_1=\prod_{a,b\geq0}\hat S_2(x+a\go_{12}-b\go_1-\go_2/2-\mu|\mu-\go_2/2,\mu+\go_2/2)\nn\\
v_2:&& I_2=\prod_{a,b\geq0}\hat S_2(x-(a+1)\go_2+b\go_{12}+\go_1/2-\mu|\mu-\go_1/2,\mu+\go_1/2)^{-1}\nn\\
v_3:&& I_3=\prod_{a,b\geq0}\hat S_2(x-(a+1)\go_1-b\go_2+\go_{12}/2-\mu|
\mu+\go_{12}/2,\mu-\go_{12}/2)^{-1}\nn\eea
where $\go_{12}$ is short for $\go_1-\go_2$.

One can likewise check the location and multiplicity of zeros and poles of the product of the above three factors tally exactly with fig.\ref{fig:PolygonPointsSwann}. For example, at $0\go_1+0\go_2+4\mu$, factor $I_1$ alone contributes $(x+4\mu)^3$, agreeing with the multiplicity 3 from fig.\ref{fig:PolygonPointsSwann}. In contrast, at $2\go_1-\go_2+4\mu$
factor $I_1$ contributes $(x+2\go_1-\go_2+4\mu)^2$ at $(a,b)=(2,0)$ or $(3,1)$, while $I_2$ gives $(x+2\go_1-\go_2+4\mu)^{-1}$ at $a=0=b$. Finally at $-3\go_1-\go_2+4\mu$, we have that $I_1$ and $I_3$ contribute multiplicity $+4$ and $-4$ and cancel out. This check shows that up to Bernoulli we have factorisation
\bea S^{C(X)}(x|\go_1,\go_2,\mu)\sim I_1 I_2 I_3.\nn\eea

\section{Factorisation in the General case}\label{sec_FitGa}
In this section, we give only a recipe while the proofs will be relegated to \cite{HTaSFI}. But we sketch the idea here. 

From the equivariant localisation of index calculation, one gets from each ${\mathbb T}^2$ fixed locus a contribution that we call a fractional $S_2$ function. For the case of $S^7$ or Swann bundle, the fractional $S_2$ is just the usual $S_2$ or the half-$S_2$ defined in \eqref{eq:S2halfdef}. And just as in those cases, the under/over-counting will cancel once the contribution from all fixed loci are included (this follows from index theorem). This means that the $S^{C(X)}$ has matching poles and zeros with the product of fractional $S_2$'s from all fixed loci. The pain lies in figuring out the Bernoulli factors. We do so by proving an auxiliary  factorisation result, and this allows us to get the asymptotic behaviour of \eqref{eq:S2halfdef} and as the asymptotics are controlled by the Bernoulli, we can by comparison figure out the missing Bernoulli factors.

The mathematical proof does not require knowledge of the local geometry at torus fixed loci, but such knowledge does offer a reassuring check.

\subsection{Fractional $S_2$ function}
Each ${\mathbb T}^2$ fixed locus result in a 'fractional' $S_2$ function
\bea &&S_2(x|p,r|\vec\go)=e^{-i\pi B}\prod_{\vec m\in I}(x+\vec m\cdotp\vec\go)
\prod_{\vec m\in II}(-x+\vec m\cdotp\vec\go)^{-1},\label{frac_S2}\\
&&I=\{\vec m=(a,b)|a,b\geq0,~r|(a-b-p-1)\}\nn\\
&&II=\{\vec m=(a,b)|a,b>0,~r|(a-b-p-1)\}.\nn\eea
If we set $r=2$ and $p$ is even, we get back the half-$S_2$ above. 
The Bernoulli term $B$ reads
\bea B(x)=\frac{x^2}{r\go_1\go_2}-\frac{(\go_1+\go_2)x}{r\go_1\go_2}+
\frac{1}{6r\go_1\go_2}(\go_1^2+\go_2^2+\go_1\go_2(6(p+1)^2-6(p+1)r+r^2+2)).\nn\eea
The $B$ factor is chosen so that $S_2$ has trivial asympotitic behaviour
\bea \lim_{\im x\to+\infty}S_2(x|p,r,\vec\go)=1.\label{triv_asymp_S2}\eea
Setting $r=1$, the $B(x)$ reverts to the standard Bernoulli polynomial $B_{2,2}(x|\vec\go)$ defined in \cite{Narukawa}.

Denote with $X_i$ the neighbourhood of the $i^{th}$ ${\mathbb T}^2$ fixed locus associated with the vector $v_i$. To get the correct fractional $S_2$ from $X_i$, we define a few more quantities.
We pick for each $v_i$ the integer vector $\hat v^i$ such that $\hat v^i\cdotp v_i=1$ and let $q_i=\hat v^i\cdotp P_i$, $r_i=v_i\times P_i$. Also pick an arbitrary real 2-vector $\vec\ep=[\ep_1,\ep_2]$, define 
\bea &&\xi_{i1}=-\frac{1}{v_i\times P_i}P\times[\go_1,\go_2],~~
\xi_{i2}=\frac{1}{v_i\times P_i}(P_i-2v_i)\times[\go_1,\go_2],\nn\\
&&\ep_{i1}=-\frac{1}{v_i\times P_i}P\times[\ep_1,\ep_2],~~
\ep_{i2}=\frac{1}{v_i\times P_i}(P_i-2v_i)\times[\ep_1,\ep_2],\nn\\
&&[m_i]=\Big\{\begin{array}{cc}m_i & \ep_{i1}<0 \\
-m_i-1 & \ep_{i1}>0  \end{array},~~~[n_i]=\Big\{\begin{array}{cc}n_i & \ep_{i2}<0 \\
-n_i-1 & \ep_{i2}>0  \end{array},\nn\\
&&s_{i1}=\Big\{\begin{array}{cc}1 & \ep_{i1}<0 \\
-1 & \ep_{i1}>0  \end{array},~~~s_{i2}=\Big\{\begin{array}{cc}1 & \ep_{i2}<0 \\
-1 & \ep_{i2}>0  \end{array},\nn\\
&&p_i=([m_i]-[n_i])q_i+2[n_i].\label{adp_weights}\eea
With these definitions the $X_i$ will contribute
\bea I_i&=&\prod_{m_i,n_i\geq0}S_2(x+([m_i]+1/2)\xi_{i1}+([n_i]+1/2)\xi_{i2}-\mu|\nn\\
&&\hspace{3cm}|p_i,r_i|\mu-(\xi_{i1}+\xi_{i2})/2,\mu+(\xi_{i1}+\xi_{i2})/2)^{s_{i1}s_{i2}}.\label{factor_i}\eea
Note $q_i$ is only well-defined up to multiples of $r_i$, but this is not going to affect summation ranges I, II in \eqref{frac_S2} and so the fractional $S_2$ function is well-defined.

\subsection{The factorisation theorem}\label{sec.Tft}
Following the notation in sec.\ref{sec:counting} in particular sec.\ref{sec:Hfitolpiac}, given the $\tilde Q$ matrix, whose rows are the vectors $v_i$, we compute $\vec N^{\pm}_i$ as in \eqref{eq:Nidef}. These vectors are the normal to a 3D cone $C$. The special function $S^C(x|\go_1,\go_2,\mu)$ defined in \eqref{eq:SQdef} associated to the cone has the factorisation:
\bea S^{C(X)}=e^{i\pi B}\prod_iI_i,\label{fac_theorem}\eea
where $I_i$ is defined in \eqref{factor_i}. 
Note the choice of $\vec\ep$ in defining $I_i$ will drop out after taking the product.

The Bernoulli factor is as follows. Consider the sum
\bea e^{xt}\sum_{\vec m}\exp(\vec m\cdotp (\go_1,\go_2,\mu)t) \nn\eea
where the sum of $\vec m$ is in the cone $C$ and with the multiplicity defined as in \eqref{eq:multiplicity}.
The Bernoulli polynomial is the coefficient of $t^0$. Note that the sum has fourth order pole at $t=0$.
The Bernoulli polynomial is of fourth order, we can only identity the $z^4$ coefficient
\bea B|_{x^4}=\frac{3}{\pi^4}\operatorname{Vol}_X=\frac{3}{2\mu}\opn{Vol}_{\Gd}.\nn\eea
Here $\opn{Vol}_X$ is the \emph{refined} volume of the 3-Sasaki manifold while $\opn{Vol}_{\Gd}$ is the volume of the cone defined in sec.\ref{sec:Hfitolpiac}, capped off at $\vec p\cdotp(\vec \go,\mu)=1$. For example for $S^7$
\bea  B|_{x^4}=\frac{3}{\pi^4}\frac{1}{(\mu^2-\go_1^2)(\mu^2-\go_2^2)}.\nn\eea
We refrain from calling our volume \emph{squashed} volume since the metric is kept 3-Sasaki. Instead, we define a 1-form $\gk$ as the unique 1-form that evaluates to 1 on the $U(1)$ vector field $R$ with weight $(\vec\go,\mu)$ and $\iota_Rd\gk=0$. We let the volume form be $\gk\wedge (d\gk)^3/24$. 

We have already sketched the proof of this theorem in the beginning of the section. 
We can establish that the two sides of \eqref{fac_theorem} have matching zeros and poles, thus we need only check the asymptotic behaviour. But we have demonstrated in \eqref{triv_asymp_S2} that the factors $I_i$ has trivial asymptotic behaviour. Thus the $B$ factor in \eqref{fac_theorem} must reproduce the asymptotic behaviour of $S^C$ function. 

To determine the asymptotic behaviour of $S^C$ function, we prove, 
by using a procedure similar to the one used in \cite{Winding:2016wpw}, another theorem which is useful in its own right: 
the function $S^C$ in \eqref{eq:SQdef} has factorisation
\bea S^C(x|\go_1,\go_2,\mu)=e^{i\pi \tilde B}\prod_i\big(\frac{x}{\tau_n}|\frac{\xi_{n1}}{\tau_n},\frac{\xi_{n2}}{\tau_n},\frac{u_n}{\tau_n}\big)_{\infty}
\big(\frac{x}{\tau_s}|\frac{\xi_{s1}}{\tau_s},\frac{\xi_{s2}}{\tau_s},\frac{u_s}{\tau_s}\big)_{\infty}\label{fac_theorem_I}\eea
where we have abused notation by using $\xi,u,\tau$ to donate the weight of the corresponding function in \eqref{tbl_wghts}, which is obtained by evaluating the weights in table \eqref{tbl_hol_fun} on a concrete set of equivariant parameters $[\vec \go,\mu]$
\bea \begin{array}{|c|c|c|}
\hline
& \textrm{north} & \textrm{south} \\
\hline 
\xi_1 & \hat v^i\cdotp (\vec \go+\mu P_i) & 2\mu+\hat v^i\cdotp (\vec \go-\mu P_i)\\
\hline
\xi_2 & 2\mu-\hat v^i\cdotp (\vec \go+\mu P_i) & \hat v^i\cdotp (-\vec \go+\mu P_i)\\
\hline
\tau & v_i\times(\vec\go+\mu P_i) & v_i\times(\vec\go-\mu P_i) \\ 
\hline u & 2\mu & 2\mu \\ 
\hline
\end{array}.\label{tbl_wghts}\eea

On the rhs the q-factorials have also trivial asymptotic behaviour. Thus $\tilde B$ must match exactly the asymptotic behaviour on the lhs. In conclusion, we see that the two Bernoulli's are equal
\bea B=\tilde B.\nn\eea

Finally we make a remark about the difference between the two factorisation theorems \eqref{fac_theorem} and \eqref{fac_theorem_I}: the second factorisation does not have an un-refined limit, that is, this formula is ill defined if
\bea \vec\go=0.\nn\eea
But the first formula remains valid, and in fact it reflects the fibration structure and the $SU(2)$ action on the 3-Sasaki manifolds.

\subsection{A final example}
We close out this section with one more example, where the integer lattice points not as easily enumerated as in the two previous cases. Consider the hyperplane arrangement with normals determined by
\begin{equation}
\Qt = \begin{pmatrix}
1 & 0 \\ 0 & 1 \\ 1 & -2 \\ 3 & 2 
\end{pmatrix} \,.   \label{eq:Qtnice4guy}
\end{equation}
The hyperplanes are illustrated in figure \ref{fig:hyperplanesnice4guy}.

We find the normals of the cone $C(X)$ as
\begin{align}                                                                                                                                                                                                                                                                                                                                                                                                                                                                                                                                                                                                                                                                                                                                                                                                                                                                                                                                                                                                                                                                                                                                                                                                                                                                                                                                                                                                                                                                                                                     \vec{N}_1^+ &= (3,5,1) \,, & \vec{N}_1^- &= (-3,-5,1)  \,,\nn \\
\vec{N}_2^+ &= (-5,1,1) \,, & \vec{N}_2^- &= (5,-1,1)  \,,  \label{eq:Nnice4guy} \\ 
\vec{N}_3^+ &= (5,1,1) \,, & \vec{N}_3^- &= (-5,-1,1)  \,, \nn \\
\vec{N}_4^+ &= (1,5,1) \,, & \vec{N}_4^- &= (-1,-5,1)  \,, \nn
\end{align}
and the  mod 2 vector $\vec{N}$ is
\begin{equation}
\vec{N} = (1,1,1) \,.
\end{equation}
To enumerate the integer lattice points in the cone described by these normals with multiplicities given by \eqref{eq:multiplicity} would be cumbersome by hand. We illustrate such lattice points in the slice $u=19$ in figure \ref{fig:PolygonPointsnice4guy}. For this example we thus leave the function $S^\Qt$ of \eqref{eq:SQdef} in its abstract form and note that we can find the first few terms with the aid of a computer.
In particular, we can compute some terms in the numerator of this function for the unsquashed case and obtain the first few terms of the Hilbert series:
\begin{equation}
1+2x^2+3x^4+4x^5+6x^6 + 8x^7 + 11 x^8 +14 x^9 +20 x^{10} + \cdots \,.
\end{equation}

\begin{figure}[tbp]
\centering
\noindent
\begin{minipage}[t]{0.45\textwidth}
\centering
\begin{tikzpicture} [scale=0.7]
\draw [->] (-5,0) -- (5,0) node [below] {\small $s$};
\draw [->] (0,-5) -- (0,5) node [left] {\small $t$};
\draw [line width=0.5ex] (-4,0) -- (4,0);
\draw [line width=0.5ex] (0,-4) -- (0,4);
\draw [line width=0.5ex] (-4,-2) -- (4,2);
\draw [line width=0.5ex] (8/3,-4) -- (-8/3,4);
\foreach \x in {-4,...,4}
\foreach \y in {-4,...,4}
	{
	\draw [black,fill=black] (\x,\y) circle (0.5ex);
	 }
\end{tikzpicture}
\caption{Hyperplanes corresponding to $\Qt$ in \eqref{eq:Qtnice4guy}.}
\label{fig:hyperplanesnice4guy}
\end{minipage}
\hfill
\noindent
\begin{minipage}[t]{0.45\textwidth}
\centering
\begin{tikzpicture}[scale=.7]
\draw [->] (-5,0) -- (5,0) node [below] {\small $s$};
\draw [->] (0,-5) -- (0,5) node [left] {\small $t$};
\draw [line width=0.5ex] (-4,0) -- (4,0);
\draw [line width=0.5ex] (0,-4) -- (0,4);
\draw [line width=0.5ex] (-4,-2) -- (4,2);
\draw [line width=0.5ex] (8/3*1.2,-4*1.2) -- (-8/3*1.2,4*1.2);
\foreach \x in {-4,...,4}
\foreach \y in {-4,...,4}
	{
	\draw [black,fill=black] (\x,\y) circle (0.5ex);
	 }
%\foreach \k in {0,2,4,6,8,10,12,14,16}
\foreach \k in {1,3,5,7,9,11,13,15,17,19}
	{\draw [line width=0.25ex, color=black] (0,\k/5)-- (2*\k/11,\k/11) --(\k/5,0)--(2*\k/13,-3*\k/13)--(0,-\k/5)--(-2*\k/11,-\k/11) -- (-\k/5,0) -- (-2*\k/13,3*\k/13)--cycle;
	}
\foreach \k in {-1,1}
	{
	\node[circle,scale=0.65,color=white, fill=black] at (0,\k) {$\boldsymbol{8}$};	
	\node[circle,scale=0.65,color=white, fill=black] at (\k,0) {$\boldsymbol{8}$};	
	\node[circle,scale=0.65,color=white, fill=black] at (0,3*\k) {$\boldsymbol{3}$};	
	\node[circle,scale=0.65,color=white, fill=black] at (3*\k,0) {$\boldsymbol{3}$};
	\node[circle,scale=0.65,color=white, fill=black] at (2*\k,1) {$\boldsymbol{5}$};
	\node[circle,scale=0.65,color=white, fill=black] at (2*\k,-1) {$\boldsymbol{5}$};
	\node[circle,scale=0.65,color=white, fill=black] at (-\k,2*\k) {$\boldsymbol{6}$};
	\node[circle,scale=0.65,color=white, fill=black] at (\k,2*\k) {$\boldsymbol{4}$};
	\node[circle,scale=0.65,color=white, fill=black] at (-2*\k,3*\k) {$\boldsymbol{4}$};
	\node[circle,scale=0.65,color=white, fill=black] at (-3*\k,2*\k) {$\boldsymbol{2}$};
	\node[circle,scale=0.65,color=white, fill=black] at (-\k,4*\k) {$\boldsymbol{1}$};
	\node[circle,scale=0.65,color=white, fill=black] at (-3*\k,4*\k) {$\boldsymbol{1}$};
	\node[circle,scale=0.65,color=white, fill=black] at (3*\k,2*\k) {$\boldsymbol{1}$};
	}
\end{tikzpicture}
\caption{Integer lattice points and their multiplicities in a slice of the cone described by the normals \eqref{eq:Nnice4guy} at $u=19$.}
\label{fig:PolygonPointsnice4guy}
\end{minipage}
\end{figure}

For the factorisation we pick
\bea \hat{v}^1 = (1,0),~~\hat{v}^2=(0,1),~~\hat{v}^3= (1,0),~~\hat{v}^4=(1,-1),~~\vec\ep = (-1,-10).\nn\eea 
We list the various quantities in \eqref{adp_weights}
\bea \begin{array}{|c|c|c|c|c|c|}
\hline
\go_{i1} & \go_{i2} & q_i & r_i & \ep_{i1} & \ep_{i2} \\
\hline
\go_1-\frac35\go_2 & -\go_1+\frac15\go_2 & 3 & 5 & - & + \\
\hline
\frac15\go_1+\go_2 & \frac15\go_1-\go_2 & 1 & 5 & + & - \\
\hline
\frac1{11}\go_1-\frac{5}{11}\go_2 & -\frac{5}{11}\go_1+\frac{3}{11}\go_2 & 5 & 11 & - & + \\
\hline
\frac{5}{13}\go_1-\frac{1}{13}\go_2 & -\frac{1}{13}\go_1-\frac{5}{13}\go_2 & -4 & 13 & - & - \\
\hline
\end{array}\nn\eea
From this table one may write down the factors for each fixed point as in \eqref{factor_i}.

\subsection{A geometric intuition and a speculation}\label{sec:Agiaas}
We give the geometrical intuition behind the rather awkward condition in the definition \eqref{frac_S2} of fractional $S_2$ that
\bea r_i~{\rm divides}~(a_i-b_i-p_i-1)\nn\eea
in the product \eqref{frac_S2}.

We stated in sec.\ref{sec_Gctatfp} that $X_i$ has local geometry
\ref{local_picture}
\bea \BB{C}^*\times_{\BB{C}^*}({\cal S}(-1)_{\BB{C}}\oplus{\cal O}(1)\oplus{\cal O}(1))\nn.\eea
Denote as we did in sec.\ref{sec_Gctatfp} with $\xi_{1,2}$ the fibre coordinate of the normal bundle. Also parametrise the $\BB{C}^2\backslash\{0\}$ (see \eqref{eq:used} for the notation) as before with $z+jw$. We can try to write the holomorphic functions by stacking monomials using $\xi_{1,2}$ and $z,w$. But not any monomial is a holomorphic \emph{function}: 
Under the $\BB{C}^*$ action the monomial $\xi_1^{[m]}\xi_2^{[n]}$ has weight $-[m_i]q_i-[n_i](2-q_i)=-p_i$, while a monomial $z^{a_i}w^{b_i}$ has then weight $a_i-b_i$.  
Note we have $a_i-b_i$ rather than $a_i+b_i$ because $w$ has opposite weight compared to $z$ under left multiplication by $\BB{C}^*$. In summary a monomial $\xi_1^{[m_i]}\xi_2^{[n_i]}z^{a_i}w^{b_i}$ has weight $a_i-b_i-p_i$. There is a final shift $a_i-b_i-p_i\mapsto a_i-b_i-p_i-1$ coming from the weight of the determinant of the normal bundle\footnote{we have already seen its effect in the usual $S_2$ function, where the product over the lattice points in the third quadrant is only over the interior. This shift is due to the canonical bundle. Now that our geometry has complex dimension 4 rather than 2, there is an extra shift compared to $S_2$.} If this weight is divisible by $r_i$, then the monomial can be regarded as a holomorphic function. This way we have explained the awkward product  rule in \eqref{frac_S2}.

Our speculation concerns the representation ring structure of $SU(2)$ in the partition function, or in $S^C$. Indeed with a right $SU(2)$ action it is very tempting, in the definition \eqref{eq:SQdef}, to interpret the multiplicity as coming from $SU(2)$ representation of various spins.

However one should immediately protest such an interpretation. We computed the partition function as a super-determinant over the Dolbeault cohomology of the HK cone. But this relies on a choice of the complex structure, which will be scrambled by the right $SU(2)$ action, in particular, there is no $SU(2)$ action on the holomorphic functions. But it is also true that the original SYM theory was constructed using solely the HK or 3-Sasaki structure, without favouring any particular complex structure. It is only in the localisation computation did we fix a complex structure. So it is at least credible that the $SU(2)$ representation structure can still be present. Our proposal is as follows, had we been able to set up a localisation computation respecting the democracy of the three complex structures, then we would not have enumerated sections given in the table \eqref{wght_ex_n} \eqref{wght_ex_s}. Rather, we would enumerate those sections in the table \eqref{adp_wght_ex_n} \eqref{adp_wght_ex_s}. The sections with the same name in the two sets of tables have the same weights. But for sections of the latter two tables, the local geometry is presented in a way that respects the right $SU(2)$ action, at the price of losing holomorphy. See the discussion round \eqref{local_picture_adpt}. But surely, without holomorphy, it makes no sense to enumerate sections of a bundle. 
We suggest therefore that one needs to combine the two pictures, i.e. picture \eqref{local_picture}: $\BB{C}^*\times_{\BB{C}^*}({\cal S}(-1)_{\BB{C}}\oplus{\cal O}(1)\oplus{\cal O}(1))$ for localising and for knowing what to enumerate, and 
picture \eqref{local_picture_adpt}: ${\cal O}(-q_i)\oplus {\cal O}(q_i-2)\oplus {\cal S}(r_i)$ to make the $SU(2)$ structure explicit.

\section{Summary} \label{sec:summary}

In this paper we have studied 7D maximally supersymmetric Yang-Mills theory on 3-Sasakian manifolds. The localisation procedure in \cite{Polydorou2017} was applied to obtain the perturbative partition function in terms of holomorphic functions on the metric cone over the manifold. Restricting to the case of hypertoric 3-Sasakian manifolds, whose HK cones have hypertoric symmetry, the holomorphic functions could be described in terms of integer lattice points in a rational convex polyhedral cone determined by the hypertoric data. This result is similar in spirit to the toric Sasaki-Einstein case, where holomorphic functions are in one-to-one correspondence with integer lattice points in the moment map cone of the toric action. However, the hypertoric integer lattice count differ from the toric case in a couple of ways. Firstly,  only half of the integer lattice points in the cone contribute, namely those satisfying some (mod 2)-constraint. Secondly, the integer lattice points are counted with multiplicities, related to how far from the edge of the cone the point is.

The perturbative partition function was written in terms of a special function \eqref{eq:SQdef} that was defined in terms of infinite products over the integer lattice points in the cone. It would be interesting to study this function from a mathematical viewpoint.

In section \ref{sec:factorisation} we gave a recipe for how to factorise this function into factors involving infinite products over generalised double sine functions. We speculated that such expressions could be interpreted as perturbative Nekrasov partition functions on $\Gc\backslash S^3 \times_{\rm tw} \C^2$ and that the full partition function might be given by a similar product (up to possible fluxes).
 However, these statements are highly speculative and one would need to derive them from first principles.

Although we have obtained a closed form of the perturbative partition function for 7D SYM on a  new family of manifolds, the physical interpretation of what we have calculated is still a bit unclear. The 7D theory is non-renormalisable and so it would be natural to interpret our calculations in terms of a UV-completion of the theory. However, we do not yet know what this UV-completion is. Since 6D maximal SYM is related to `little string theory' a speculation is that the 7D theory is related to `little m-theory' \cite{Losev:1997hx}.

%%%%%%%%%%%%%%%%%%%%%%%%%%%%%%%%%%%%%%%%%%%%%%%%%%%
%%%% ACKNOWLEDGMENTS
%%%%%%%%%%%%%%%%%%%%%%%%%%%%%%%%%%%%%%%%%%%%%%%%%%%

\acknowledgments

This research is supported in part by Vetenskapsr\aa det under grant \#2014-5517, by the STINT grant and by the grant ``Geometry and Physics'' from the Knut and Alice Wallenberg foundation.

%%%%%%%%%%%%%%%%%%%%%%%%%%%%%%%%%%%%%%%%%%%%%%%%%%%
%%%% APPENDICES
%%%%%%%%%%%%%%%%%%%%%%%%%%%%%%%%%%%%%%%%%%%%%%%%%%%

%%%%%%%%%%%%%%%%%%%%%%%%%%%%%%%%%%%%%%%%%%%%%%%%%%%
%%%% REFERENCES
%%%%%%%%%%%%%%%%%%%%%%%%%%%%%%%%%%%%%%%%%%%%%%%%%%%

%\bibliographystyle{JHEP}
%\bibliography{PhysicsPaperBib}{}
\providecommand{\href}[2]{#2}\begingroup\raggedright\endgroup

\end{document}